\documentclass[onefignum,onetabnum]{siamart190516}

\usepackage[utf8]{inputenc}

\usepackage{lineno}
\modulolinenumbers[5]

\usepackage{physics, siunitx, graphicx, amsmath, amssymb, mathtools, amsfonts, pgfplots, caption, subcaption, epstopdf, booktabs}
\usepackage[version=4]{mhchem}
\usepackage{comment}
\usepackage{placeins}
\usepackage{tikz}
\usetikzlibrary{patterns,arrows}
\pgfplotsset{compat=1.13}
\usepackage{todonotes}
\usepackage{overpic}

\usepackage{collcell}
\usepackage{etoolbox}

\newtoggle{inTableHeader}
\toggletrue{inTableHeader}
\newcommand*{\StartTableHeader}{\global\toggletrue{inTableHeader}}%
\newcommand*{\EndTableHeader}{\global\togglefalse{inTableHeader}}%

\let\OldTabular\tabular%
\let\OldEndTabular\endtabular%
\renewenvironment{tabular}{\StartTableHeader\OldTabular}{\OldEndTabular\StartTableHeader}%
\newcommand*{\MinNumber}{0.0}%
\newcommand*{\MidNumber}{0.5} %
\newcommand*{\MaxNumber}{1.0}%

\newcommand{\ApplyGradient}[1]{%
  \iftoggle{inTableHeader}{#1}{
    \ifdim #1 pt > \MidNumber pt
        \pgfmathsetmacro{\PercentColor}{max(min(100.0*(#1 - \MidNumber)/(\MaxNumber-\MidNumber),100.0),0.00)} %
        \hspace{-0.33em}\pgfsetfillopacity{0.35}\colorbox{red!\PercentColor!blue}{\pgfsetfillopacity{1}#1}
    \else
        \pgfmathsetmacro{\PercentColor}{max(min(100.0*(\MidNumber - #1)/(\MidNumber-\MinNumber),100.0),0.00)} %
        \hspace{-0.33em}\pgfsetfillopacity{0.35}\colorbox{blue!\PercentColor!red}{\pgfsetfillopacity{1}#1}
    \fi
  }}
\newcolumntype{R}{>{\collectcell\ApplyGradient}c<{\endcollectcell}}

\newcommand{\rob}[1]{\todo[inline,caption={},color=red!40]{Rob: #1}}

\newcommand{\sep}{\mu}
\newcommand{\aone}{\ell_1}
\newcommand{\atwo}{\ell_2}
\newcommand{\del}{\nabla}

\newcommand{\non}{\nonumber}

\newcommand{\veps}{\varepsilon}
\newcommand{\beqa}{\begin{eqnarray}}
\newcommand{\eeqa}{\end{eqnarray}}
\newcommand{\beqas}{\begin{eqnarray*}}
\newcommand{\eeqas}{\end{eqnarray*}}
\newcommand{\beq}{\begin{equation}}
\newcommand{\eeq}{\end{equation}}

\newcommand{\pdhfrac}[2]{\mathchoice{\frac{#1}{#2}}{#1/#2}{#1/#2}{#1/#2}}
\newcommand{\fdd}[2]{\pdhfrac{\mathrm{d}#1}{\mathrm{d}#2}}
\newcommand{\sdd}[2]{\pdhfrac{\mathrm{d}^2#1}{\mathrm{d}#2^2}}
\newcommand{\pd}[2]{\pdhfrac{{\partial}#1}{{\partial}#2}}
\newcommand{\spd}[2]{\pdhfrac{\partial^2#1}{{\partial}#2^2}}
\newcommand{\mpd}[3]{\pdhfrac{\partial^2#1}{{\partial}#2{\partial}#3}}
\newcommand{\tdd}[2]{\pdhfrac{\mathrm{d}^3#1}{{\mathrm{d}}#2^3}}

\renewcommand{\d}[1]{\mathrm{d}#1}
\newcommand{\ra}{\rightarrow}
\newcommand{\eref}[1]{(\ref{#1})}
\newcommand{\Fref}[1]{Figure~\ref{#1}}

\newcommand{\PDD}[2]{\frac{\partial^2{#1}} {\partial{#2^2}}}

\usepackage{amsfonts}
\usepackage{graphicx}
\usepackage{epstopdf}
\usepackage{algorithmic}
\ifpdf
  \DeclareGraphicsExtensions{.eps,.pdf,.png,.jpg}
\else
  \DeclareGraphicsExtensions{.eps}
\fi

\headers{Spirally-wound high-contrast layered materials}{S. Psaltis, R. Timms, C.P. Please and  S.J. Chapman}

\title{Homogenisation of spirally-wound high-contrast layered materials
  \thanks{Submitted to the editors 9 November, 2020.
\funding{This publication is based on work supported with funding provided by The Faraday Institution, grant number EP/S003053/1, FIRG003, and by the Australian Research Council Centre of Excellence for Mathematical and Statistical Frontiers (project number CE140100049), funded by the Australian Government.}}}

\author{Steven Psaltis\footnotemark[2]
  \and Robert Timms\footnotemark[3] \footnotemark[4]
  \and Colin Please\footnotemark[3] \footnotemark[4]
\and S.~Jonathan Chapman\footnotemark[3] \footnotemark[4]}

\usepackage{amsopn}

\ifpdf
\hypersetup{
  pdftitle={Homogenisation of spirally-wound high-contrast layered materials},
  pdfauthor={S. Psaltis, R. Timms, C.P. Please and  S.J. Chapman}
}
\fi
\begin{document}

\maketitle

\renewcommand{\thefootnote}{\fnsymbol{footnote}}
\footnotetext[2]{School of Mathematical Sciences, Queensland University of Technology, Brisbane QLD 4000, Australia and ARC Centre of Excellence for Mathematical and Statistical Frontiers, Queensland University of Technology, Brisbane QLD 4000, Australia (\email{steven.psaltis@qut.edu.au}).}
\footnotetext[3]{Mathematical Institute, Andrew Wiles Building, University of Oxford, Woodstock Road, Oxford OX2 6GG, UK (\email{timms@maths.ox.ac.uk}, \email{please@maths.ox.ac.uk}, \email{chapman@maths.ox.ac.uk}).}
\footnotetext[4]{The Faraday Institution, Quad One, Becquerel Avenue, Harwell Campus, Didcot, OX11 0RA, UK.}
\renewcommand{\thefootnote}{\arabic{footnote}}

\begin{abstract}
  Asymptotic homogenisation is used to systematically derive reduced-order macroscopic models of conductive behaviour in spirally-wound layered materials in which the layers have very different conductivities. The problem is motivated by the need for simplified models of the electrical and thermal behaviour of lithium-ion cells, accounting for the highly conductive metallic current collectors and relatively poorly conductive electrodes. We identify and study three distinguished limits, and then describe two composite models which each provide a uniform approximation spanning two distinguished limits.
 We compare the results of the various reduced-order models with calculations of the full model on a detailed geometry to give a guide to the accuracy of the approximations. 
\end{abstract}



\section{Introduction} 
\label{sec:intro}
Lithium-ion batteries have become one of the most common storage devices for electrical energy, with applications ranging from mobile devices to electric vehicles \cite{VanNoorden2014, scrosati2010, armand2008, zubi2018}. Mathematical models of such batteries are commonly used to  understand better their behaviour and improve their design. In order to be practical and useful, models must account for a range of physical processes occurring within the battery while remaining sufficiently computationally efficient. For some situations, simple equivalent circuit models are adequate \cite{madani2018}, while others require partial differential equations (PDEs) describing electrochemical, thermal and mechanical behaviour.  More complicated PDE models are often simplified via asymptotic analysis \cite{marquis2019, richardson2019, moyles2018}, creating a  range  of  models  of  different complexity and fidelity that account for different physical mechanisms. One major barrier to obtaining efficient numerical solutions is that the geometry of typical cells is quite complicated. A common approach is to consider a one-dimensional model in the through-cell direction, justified by the small aspect ratio of typical cell designs \cite{timms2020, marquis2020}. The standard through-cell model that accounts for the detailed porous media effects within the battery is the so-called ``pseudo two-dimensional" porous electrode model of Doyle, Fuller and Newman (DFN)\cite{Doyle1993}. The DFN, and similar models, may be derived systematically from equations posed at the microscale using volume-averaging or homogenisation techniques \cite{hunt2020, richardson2012, moyles2018}. What has received less attention is the problem of including the effects of the
larger-scale geometry of battery cells. Effects on this scale are of particular importance in larger-sized batteries, such as those used in the electric vehicle sector, which can exhibit non-uniform behaviour in the current and temperature distribution, leading to a decrease in performance and lifetime  \cite{kosch2018, rieger2016}.  

Lithium-ion cells comprise five main components: a negative current collector, negative electrode, separator,  positive electrode, and positive current collector. The current collectors in a cell are typically made of highly electrically and thermally conductive materials, such as copper or aluminium, with the two active electrode regions and the separator sandwiched between them. The distances between the current collectors is small, around 250 $\mu$m, and, in order to get sufficient active material into the battery, the layers of current collectors and electrodes are typically either layered (to form a pouch cell), folded (to form a prismatic cell), or wound (to form a cylindrical or ``jelly roll'' cell). An overview of battery construction can be found in \cite{plett2015battery}. Accounting for this complex geometry while describing behaviour of the active electrodes and separator using a model from the existing hierarchy of through-cell models is a difficult computational task. Motivated by these practical problems our aim is to present a systematic study of simplifications to idealised problems with large rapid spatial variations in conductivity. In doing so, we derive models on the macroscopic lengthscale which avoid the computational complexity of the detailed geometry, but accurately account for it. Although we begin from a problem motivated by batteries, the techniques developed are general and can be applied to many other physical systems.

\begin{figure}
    \centering
    \includegraphics[width = 0.85\textwidth, trim = {0 10cm 0 5cm}, clip = true]{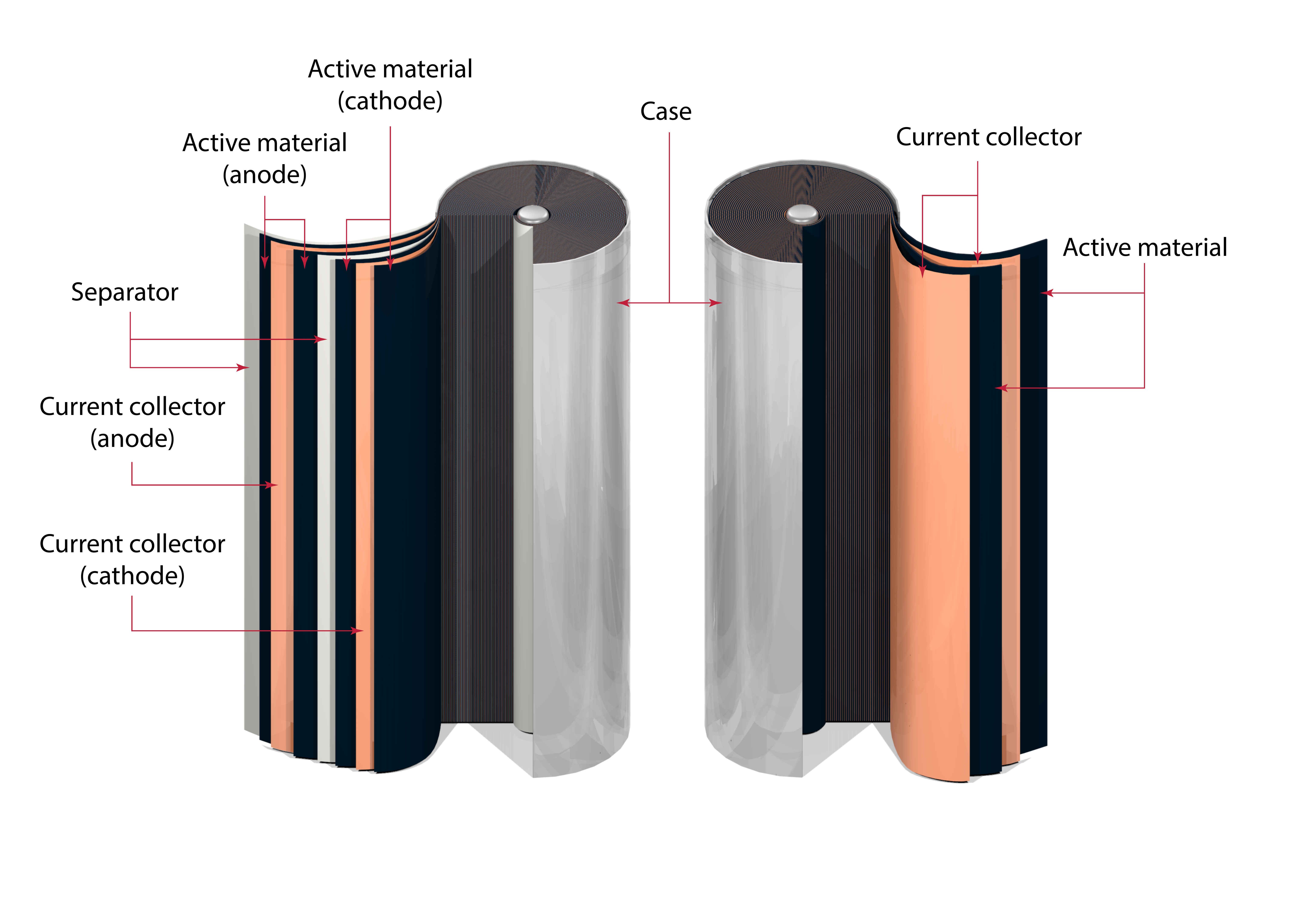}
    \caption{Simplified schematic of a ``jelly roll'' cell, showing the main components. In this work we combine the anode and cathode active regions and the separator into ``active material'' regions (right).}
    \label{fig:Jelly}
\end{figure}

In the literature there have been a number of computational approaches which account for the complicated spiral geometry when modelling the behaviour of cylindrical cells \cite{Evans1989, Saw2013, Shi2018, tranter2020}. Setting up a mesh that accounts for the structure of the spiral is a considerable problem, and the typical approach is to ``unwind" the battery and consider a two-dimensional problem along the length of the current collectors. Examples of this approach have proved effective in describing electrical behaviour and include \cite{Harb1999,Chen2006,Lee2013,McCleary2013,Guo2014,Duan2018}. Of particular note is the approach of Guo \& White \cite{Guo2014}. In their work, the electrical problem in the unrolled geometry is treated using the ``two-potential'' approach, in which local problems for the behaviour of the active part of the cell (cathode, separator, anode) are coupled via an electrical problem that considers a potential for each current collector. The thermal problem is then solved on the original cylindrical geometry with the heating predicted by the two-potential model averaged over different cell regions and with anisotropic thermal diffusion. Our approach here will demonstrate that elements of this are natural simplifications in the parameter regimes of interest, but that these approximations can be derived systematically and further simplifications can be made without degrading the accuracy of the approximation. 

We focus our attention on the macroscopic geometry of a spirally wound cell, where the microstructure can be represented by a series of layers. We treat a general problem, described in  Section~\ref{sec:Model}, which encompasses both the electrical and thermal problem in the cell in the simplest case of all materials behaving as Ohmic resistors. Extending this description to more realistic battery models is straightforward, but with the resulting computational overhead. In Section~\ref{sec:homogenisation}, through a multiple-scales homogenisation, we derive a set of models governing the potential throughout the cell depending on the relative conductivity of each of the materials.
 The main outcome of the analysis is that there are three distinguished limits for the conductivity ratio relative to the lengthscale ratio.  These correspond to a conventional homogenised model in which there is an effective diffusion coefficient, a novel limit of enhanced radial diffusion, and a ``two-potential'' limit. We discuss each of these in turn and indicate where existing approaches to macroscopic modelling of batteries fits within these cases.

 Each of the models is reviewed in Section \ref{sec:review}, and practical matters, such as composite expansions and extensions to other geometries
  are discussed. In Section~\ref{sec:results} the results from the homogenised models are compared to the solution obtained from directly solving the governing equations on the spiral geometry in COMSOL. Finally, in Section~\ref{sec:conclusion} we present our conclusions.

\section{The model}
\label{sec:Model}

\subsection{Geometry}
\label{sec:geometry}
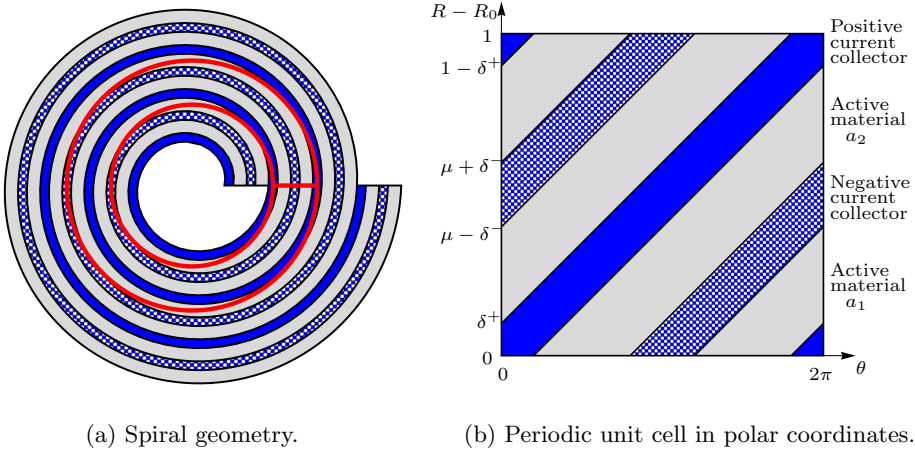
\begin{figure}[htbp]
  \centering
  \begin{subfigure}[t]{0.45\textwidth}
 \includegraphics[width=\textwidth]{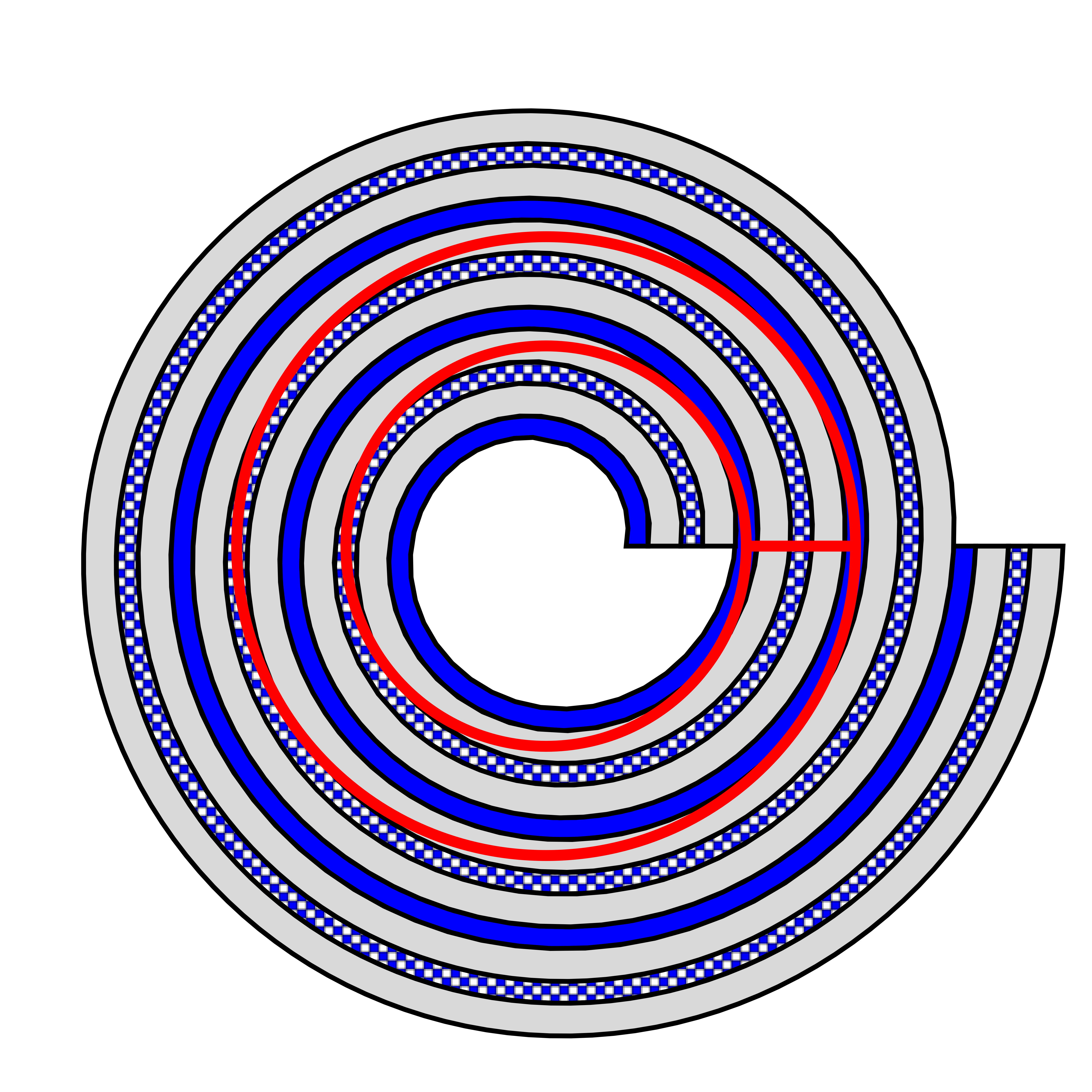}
  \caption{Spiral geometry.}
  \label{fig:spiral_polar}
  \end{subfigure}%
~
  \begin{subfigure}[t]{0.53\textwidth}
    \begin{overpic}[width=\textwidth, trim={-2.2cm  -1.5cm -3cm 0}, clip=true]{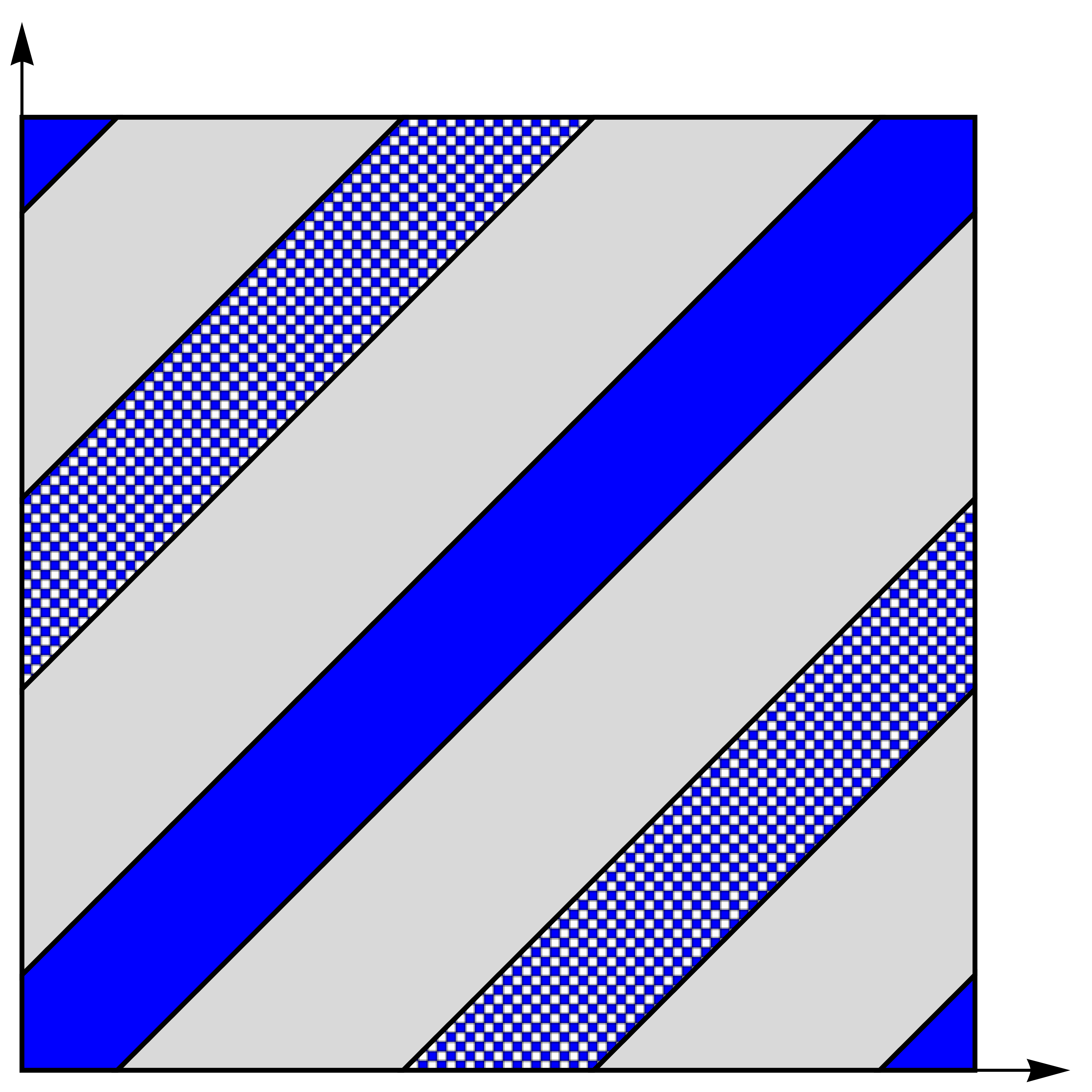}
      \scriptsize
      \put(82,6){$\theta$}
      \put(73,5){$2 \pi$}
      \put(13,5){$0$}
      \put(10,8){$0$}
      \put(9,15){$\delta^+$}
      \put(2,32){$\sep-\delta^-$}
      \put(2,45){$\sep+\delta^-$}
      \put(2,64){$1-\delta^+$}
      \put(10,70){$1$}
\put(0,75){$R-R_0$}
      \put(77,72){Positive}
      \put(77,69){current}
      \put(77,66){collector}
      \put(77,57){Active}
      \put(77,54){material}
      \put(80,51){$a_2$}
      \put(77,42){Negative}
      \put(77,39){current}
      \put(77,36){collector}
      \put(77,25){Active}
      \put(77,22){material}
      \put(80,19){$a_1$}
      \end{overpic}
\caption{Periodic unit cell in  polar coordinates.}
\label{fig:cell_polar}
\end{subfigure}
\caption{Representation of spiral geometry. The left figure shows the original geometry, with the periodic unit cell outlined in red. The right figure shows the same periodic cell in dimensionless  polar coordinates $R = r/\varepsilon$ and $\theta$.  }
\label{fig:spiral}
\end{figure}

We consider a spirally-wound cell, such as the common 18650 lithium-ion cell, where the geometry through a cross-section is illustrated in \Fref{fig:spiral_polar}. In practice these cells are constructed by rolling a sandwich of layers containing the active cathode, positive current collector, active cathode, separator, active anode, negative current collector, active anode, and separator. However we ignore the details of the anode, cathode and separator, and treat them as a single region of active material, modelled as an Ohmic conductor, with two such regions per winding.
We assume that the current collectors and the active layers all have uniform thickness and the total thickness of the sandwich of the repeating layers is $h$. 
We take the positive and negative current collectors to have thickness $2\delta^{+} h$ and $2\delta^{-} h$ respectively
(with $0<\delta^{\pm}<1/4$),
and the active layers to have thickness 
$\aone h$ and $\atwo h$, with $\aone+\atwo+2\delta^++2\delta^-=1$.
We take the inner radius of the cell to be $L_0$, and the outer radius to be $L$.

Assuming the cell is rolled as an Archimedean spiral, as shown in \Fref{fig:spiral_polar}, and using polar coordinates, the boundaries of the positive current collector are given by $r = L_0 \pm \delta^+ h + h \theta/2\pi$, and  the boundaries of the negative current collector are given by $r = L_0 +  \sep h \pm \delta^- h  + h \theta/2\pi$, where $\aone = \sep-\delta^--\delta^+$ and $\atwo =1-\sep-\delta^--\delta^+$.

\subsection{Physics}

We distinguish the variables/parameters in  different material regions using  the superscript $i$, with \mbox{$i\in\{+, -, a_1,a_2\}$,} representing the positive current collector, the negative current collector, and the two active material regions, respectively. The current density, $\vec{i}^i$, in each region can be written as
\begin{equation}
    \vec{i}^{i} = -\sigma^{i} \grad \phi^{i},
\end{equation}
where $\phi^{i}$ is the electric potential, and $\sigma^{i}$ is the conductivity. The electric potential satisfies 
\begin{equation}
    \grad  \cdot \, (\sigma^{i} \grad \phi^{i}) = 0. \label{eq:potential}
\end{equation}
At interfaces between the materials the component of the current density in the normal direction must be continuous, and the potential must also be continuous. 

For the thermal problem we assume heat is transported by diffusion only and generated by Ohmic (or Joule) heating due to the passage of current. In the interest of simplicity we consider the steady problem and ignore any transient effects, though these can be accounted for in a straightforward manner.
The thermal problem is therefore governed by
\begin{equation}
    \grad \cdot \left(k^i \grad T^i \right) + \sigma^i |\grad\phi^i|^2 = 0,
    \label{eq:Thermal}
\end{equation}
where $k^i$ is the thermal conductivity in region $i$. Again, at interfaces between materials the component of the heat flux $k^i \grad T^i$ in the normal direction must be continuous, and the temperature must also be continuous.

We may treat both (\ref{eq:potential}) and (\ref{eq:Thermal}) with a single analysis by solving the general problem
\begin{equation}
 -  \grad  \cdot \, (\sigma^{i} \grad \phi^{i}) = F^i(\vec{x}).
\label{eqn:problem}
\end{equation}
where $F^i(\vec{x})$ is given, which encompasses both models.

Finally, the problem to be solved requires boundary conditions. For the electrical problem we specify the potential at the tabs (the terminals where the battery is connected to an external circuit). In the simplest case there is a single tab on each current collector where different potentials can be applied. On the remaining parts of the boundary there is no normal component of current. The thermal boundary conditions may also be given on different parts of the boundary, and in most cases the external boundary will allow some heat flux and the tabs may have some different heat transfer behaviour.
In the interests of simplicity we suppose that these external boundary conditions allow a two-dimensional solution, so that $\phi$ is independent of position along the axis of the cell.

\subsection{Nondimensionalisation}
\label{sec:nondim}

We use $L$ as the unit of length, so that the nondimensional radius of the cell is equal to unity, and use $\sigma^+$ as the reference conductivity. The source term $F$ is scaled relative to a typical value, $[F]$, say. The appropriate scale for the potential $\phi$ depends in a not-completely-trivial way on the relative size of the conductivities $\sigma^i$. We begin by using $[F] L^2/\sigma^+$ as the scale, and will rescale when appropriate (if $F$ is zero then the scale for the potential is set by the applied voltage, which is equivalent to taking $[F]=\sigma^+ V/L^2$). Thus we set
\[ \phi^i = \frac{[F] L^2}{\sigma^+} \phi^i_*, \qquad
r = L r_*, \qquad F^i = [F] F^i_*, \qquad \sigma^i = \sigma^+ \sigma^i_*.\] 
Dropping the star notation and using dimensionless variables  henceforth
the boundaries of the positive and negative current collector are now given by $r = r_0 \pm \varepsilon \delta^+ + \varepsilon \theta/2\pi$ and
$r = r_0 + \varepsilon \sep \pm \varepsilon \delta^- + \varepsilon \theta/2\pi$, respectively, where $r_0  = L_0/L$ and $\varepsilon=h/L \ll 1$.
Equation (\ref{eqn:problem}) becomes
\begin{align}
	\frac{\sigma^i}{r}\pd{}{r}\left(r\pd{\phi^i}{r}\right) + \frac{\sigma^i}{r^2}\spd{\phi^i}{\theta} = -F^i,\quad i \in \{+, -, a_1, a_2\},\label{eq:main}
\end{align}
with boundary conditions
\begin{align}
	\phi^+ &= 
	\begin{cases}
		\phi^{a_1} & \text{on} \ r = r_0 + \varepsilon\left(\delta^+ + \frac{\theta}{2\pi}\right)	\label{eq:nondimBndconds1}\\
		\phi^{a_2} & \text{on} \ r = r_0 + \varepsilon\left(1-\delta^+  + \frac{\theta}{2\pi}\right)		
	\end{cases}\\
	\phi^- &= 
	\begin{cases}
	\phi^{a_1} & \text{on} \ r = r_0 + \varepsilon\left(\sep - \delta^- + \frac{\theta}{2\pi}\right)\\
	\phi^{a_2} & \text{on} \ r = r_0 + \varepsilon\left(\sep + \delta^- + \frac{\theta}{2\pi}\right)	
	\end{cases}	\label{eq:nondimBndconds2}\\
\sigma^+	\vec{n} \cdot \del \phi^+ &= 
	\begin{cases}
		\sigma^{a_1} \vec{n} \cdot \del\phi^{a_1} & \text{on} \ r = r_0 + \varepsilon\left(\delta^+ + \frac{\theta}{2\pi}\right)\\
		\sigma^{a_2} \vec{n} \cdot \del\phi^{a_2} & \text{on} \ r = r_0 + \varepsilon\left(1-\delta^+ + \frac{\theta}{2\pi}\right)
	\end{cases}\label{eq:nondimBndconds3}\\
\sigma^-	\vec{n} \cdot \del\phi^- &= 
	\begin{cases}
		\sigma^{a_1} \vec{n} \cdot \del\phi^{a_1} & \text{on} \  r = r_0 + \varepsilon\left(\sep - \delta^- + \frac{\theta}{2\pi}\right)	\\
		\sigma^{a_2} \vec{n} \cdot \del\phi^{a_2} & \text{on} \  r = r_0 + \varepsilon\left(\sep + \delta^- + \frac{\theta}{2\pi}\right),
	\end{cases}
	\label{eq:nondimBndconds4}
\end{align}
where $\vec{n}$ is the unit normal to the interface.
Note that we find it useful to continue to write $\sigma^+$, even though our nondimensionalisation implies it is equal to unity.

To evaluate the normal derivatives in (\ref{eq:nondimBndconds3})-(\ref{eq:nondimBndconds4}), we note that if the interface is given by the equation
\begin{equation}
f(r,\theta)=r-\frac{\varepsilon \theta}{2 \pi} = \text{constant},
\end{equation}
then we can write 
\begin{align}
\label{eqn:normal}\vec{n}\cdot\nabla \phi = \frac{\nabla f \cdot \nabla \phi}{|\nabla f|} 
& =
\left(\pd{f}{r}\pd{\phi}{r} +\frac{1}{r^2}\pd{f}{\theta}\pd{\phi}{\theta} \right)\left(\left(\pd{f}{r} \right)^2 +\frac{1}{r^2}\left(\pd{f}{\theta} \right)^2  \right)^{-1/2}
 \\ &
= \left(\pd{\phi}{r}-\frac{\varepsilon}{2 \pi r^2}\pd{\phi}{\theta} \right)\left(1+\frac{\varepsilon^2}{4 \pi^2r^2} \right)^{-1/2}.
\nonumber
\end{align}

We aim to solve (\ref{eq:main})-(\ref{eqn:normal}) in the limit $\varepsilon \ra 0$, with the parameters $\delta^{\pm}$, $\sep$ and $r_0$ taken to be $O(1)$, since this is typically the case.
However, in practical battery designs the current collector materials are much more conductive than the active material, both in terms of electrical and thermal conductivity, so that $\sigma^{a_1} \sim \sigma^{a_2} \ll \sigma^- \sim \sigma^+ = 1$. Thus in taking the limit $\varepsilon \ra 0$ it is crucial to determine how the various conductivities should scale with $\varepsilon$.

We will find that there are in fact three distinguished limits: $\sigma^{a_{1,2}}=O(\sigma^{\pm})$, termed poorly conductive current collectors; $\sigma^{a_{1,2}}=O(\varepsilon^2\sigma^{\pm})$, termed reasonably conductive current collectors; and $\sigma^{a_{1,2}}=O(\varepsilon^4\sigma^{\pm})$, termed very conductive current collectors. The poorly conductive limit is presented first and provides the result typically obtained from standard homogenisation theory. The very conductive case has an  interesting microscale structure, and hence is presented second. 
The intermediate case, which has elements from both the earlier examples, is given in the supplementary material.
 Details of the analysis are given in Section~\ref{sec:homogenisation}. Readers interested only in the resulting models are directed to the summary provided in Section~\ref{sec:review}.

\section{Homogenisation of a rolled cell}
\label{sec:homogenisation}

We exploit the method of multiple-scales homogenisation, writing the problem so that the solution depends on both macroscale and microscale variables \cite{pavliotis2008, sanchez1980,papanicolau1978, davit2013}. Crucially our geometry is periodic in polar coordinates: increasing $\theta$ by $2 \pi$ is equivalent to increasing $r$ by $\varepsilon$, as illustrated in \Fref{fig:cell_polar}. This enables us to follow the usual procedure of imposing 
periodicity on the microscale, with a slow modulation on the macroscale.
 The periodically repeated unit cell in cylindrical coordinates corresponds to an annular region bounded by concentric circles, as indicated by the boxed region in \Fref{fig:spiral_polar}. 

Introducing the microscale variable $R = r/\varepsilon$ we write $\phi = \phi(r,R,\theta)$ and treat $R$ and $r$ as independent variables, eliminating the degeneracy that this introduces by requiring that the solution is exactly periodic in $R$ with unit period. 

In the following we assume that the source term takes the form $F=F(r, R-\theta/2\pi)$ with $F$ periodic in $R-\theta/2\pi$ with period 1, which is true for the physical problems we are considering.

\subsection{Poorly conductive current collectors}
We begin with the standard homogenisation problem in which each $\sigma^i = O(1)$ for all $i$. 

After introducing $R$, equation (\ref{eq:main}) becomes
\begin{equation}
  \frac{1}{r} \left(\pd{}{r} + \frac{1}{\varepsilon}\pd{}{R}\right)
\left(  \sigma(R-\theta/2 \pi) r \left(\pd{\phi}{r} + \frac{1}{\varepsilon}\pd{\phi}{R}\right)\right) + \frac{1}{r^2}\spd{\phi}{\theta} = -F(r,R-\theta/2\pi),
  \end{equation}
where
\[ \sigma(y) = \begin{cases}
  \sigma^+ & -\delta^+ < y < \delta^+,\\
  \sigma^{a_1} & \delta^+ < y < \sep-\delta^-\\
  \sigma^- & \sep-\delta^- < y < \sep+\delta^-\\
  \sigma^{a_2} & \sep+\delta^- < y < 1-\delta^+
\end{cases}
\]
is periodic with unit period,  $F(r,y)$ is periodic in $y$ with unit period, and, for this section, we have dropped the superscript from $\phi$.
Expanding 
 the solution in powers of $\varepsilon$ 
\begin{equation}
\phi \sim \phi_0 + \varepsilon \phi_1 + \varepsilon^2 \phi_2 + \ldots.
\end{equation}
we find at leading order that
\begin{equation}
    \pd{}{R}\left( \sigma(R-\theta/2\pi) \pd{\phi_0}{R}\right) = 0.
    \label{eq:SHleading}
\end{equation}
Since $\phi_0$ is periodic with period $1$ in $R$ and period $2\pi$ in $\theta$, it  follows that $\phi_0$ is independent of $R$.
At the next order we find
\begin{equation}
    \pd{}{R}\left(\sigma(R-\theta/2\pi)\pd{\phi_1}{R}\right) = -\sigma'(R-\theta/2\pi)\pd{\phi_0}{r},
    \label{eq:SHfirst1}
\end{equation}
where $\sigma'(y) = \fdd{\sigma(y)}{y}$, with $\phi_1$ periodic, with period $1$ in $R$ and period $2\pi$ in $\theta$.
Integrating with respect to $R$ gives
\beq 
\pd{\phi_1}{R} = -\pd{\phi_0}{r} + \frac{A}{\sigma(R-\theta/2\pi)}.\label{phi1R}
\eeq
where $A$ is independent of $R$.
Integrating again and imposing  periodicity in $R$ gives
\[ 0 = \int_0^1\pd{\phi_1}{R} \,\d R = -\pd{\phi_0}{r} + A\int_0^1\frac{\d R}{\sigma(R-\theta/2\pi)} \qquad \Rightarrow \qquad A = \frac{1}{\int_0^1\frac{\d y}{\sigma(y)}}\pd{\phi_0}{r},\]
where we have used the fact that $\sigma$ is periodic with unit period.
At next order we find
\begin{eqnarray}
\lefteqn{    \frac{1}{r}\pd{}{r}\left(\sigma r \pd{\phi_0}{r}\right) + \frac{1}{r^2} \pd{}{\theta}\left(\sigma \pd{\phi_0}{\theta}\right) + \frac{1}{r}\pd{}{r}\left(\sigma r \pd{\phi_1}{R}\right) }&&\\
  \nonumber    && \qquad \qquad \qquad\mbox{ }+  \frac{1}{r}\pd{}{R}\left(\sigma r \pd{\phi_1}{r}\right) + \pd{}{R}\left(\sigma \pd{\phi_2}{R}\right) = -F(r,R-\theta/2\pi).
    \label{eq:SHsecond}
    \end{eqnarray}
    where $\sigma = \sigma(R-\theta/2\pi)$.
    Integrating over $R$ and imposing periodicity  gives the solvability condition 
 \[   \int_0^1   \frac{1}{r}\pd{}{r}\left(\sigma r \pd{\phi_0}{r}\right) + \frac{1}{r^2} \pd{}{\theta}\left(\sigma \pd{\phi_0}{\theta}\right) + \frac{1}{r}\pd{}{r}\left(\sigma r \pd{\phi_1}{R}\right)\, \d R=
 -\int_0^1F(r,R-\theta/2\pi)\, \d R,\]
which, on using (\ref{phi1R}), becomes
\begin{equation}
      \frac{\sigma_N}{r}\pd{}{r}\left( r \pd{\phi_0}{r}
    \right) +
    \frac{\sigma_T}{r^2} \PDD{\phi_0}{\theta} = -\bar{F} =  - \int_0^1 F(r,y)\d y,
         \label{eq:SHsolvability}
\end{equation}
where
\begin{align}
\sigma_N & =  \frac{1}{\int_0^1 \frac{\d y}{\sigma(y)}} =   
    \frac{1}{2\delta^+/\sigma^+ + 2 \delta^-/\sigma^- +\aone/\sigma^{a_1}+\atwo/\sigma^{a_2}}
  ,\\
    \sigma_T & =  \int_0^1 \sigma(y)\d y = 
    2\delta^+\sigma^{+} + 2\delta^-\sigma^{-} + \aone\sigma^{a_1} + \atwo\sigma^{a_2}.
\end{align}
The homogenised equation \eref{eq:SHsolvability} has a simple interpretation for our spirally wound layers of current collectors and active material: it says that, because the current collectors nearly form circles, in the radial direction the effective conductivity is that of the layers in series, while  in the angular direction the effective conductivity is that of the layers in parallel.

\subsubsection{Ohmic heating}
\label{sec:Ohmic}
In the physical problem we solve \eqref{eqn:problem} with $F=0$ to determine the electrical behaviour. We then use this solution to determine the source term for the thermal problem which is given by Ohmic heating. Writing this in multiple scales form and using the expansion for $\phi^i$ we evaluate this as
\beqas
 \sigma^i | \del \phi^i|^2 &=& \sigma^i \left( \frac{1}{\veps}\pd{\phi^i}{R} + \pd{\phi^i}{r}\right)^2+\frac{\sigma^i}{r^2} \left(\pd{\phi^i}{\theta}\right)^2 
\sim  \frac{\sigma_N^2}{\sigma^i} \left( \pd{\phi_0}{r}\right)^2+\frac{\sigma^i}{r^2} \left(\pd{\phi_0}{\theta}\right)^2 .
\eeqas

\subsection{Very conductive current collectors}
\label{sec:very conductive}
Having described the standard homogenisation problem, we now consider the more relevant limit in which the conductivity of the current collector materials is very large compared to that of the active material.
Specifically we consider the distinguished limit in which
$\sigma^{a_k} = \varepsilon^4 \hat{\sigma}^{a_k}$, where $\hat{\sigma}^{a_k}=O(1)$ as $\veps\ra 0$, for $k=1,2$.
We will find that, unlike in most conventional homogenisation, the lowest-order solution is not constant on the microscale, but has microscale structure, and  is modulated by not one but two functions of the slow variables (these are the potential in each of the current collectors). The homogenised equations governing these two functions come from solvability conditions that arise when we have solved the problem in the current collector region to fourth order, and in the active region to leading order.

The reduced conductivity of the active material means that the source term $F$ generates a much higher potential in this case. To maintain balance in the equations we must rescale $\phi$ with $1/\veps^2$.
After this rescaling,  \eref{eq:main} written in multiple-scales form is
\begin{align}
  \lefteqn{ \frac{\sigma^i}{r}\pd{}{r}\left(r\pd{\phi^i}{r}\right) + \frac{\sigma^i}{r^2}\spd{\phi^i}{\theta} + \frac{\sigma^i}{\varepsilon r}\pd{}{r}\left(r\pd{\phi^i}{R}\right) + \frac{\sigma^i}{\varepsilon r}\pd{}{R}\left(r\pd{\phi^i}{r}\right) + \frac{\sigma^i}{\varepsilon^2 }\spd{\phi}{R} } \hspace{4cm}&\label{eqn:nondimLaplacepm} \\
  &= -\veps^2 F^i(r,R-\theta/2\pi),\quad i \in \{+, -\}
\nonumber		,\\[0mm]
\lefteqn{
  \frac{\veps^4\hat{\sigma}^i}{r}\pd{}{r}\left(r\pd{\phi^i}{r}\right) + \frac{\veps^4\hat{\sigma}^i}{r^2}\spd{\phi^i}{\theta} + \frac{\veps^3\hat{\sigma}^i}{ r}\pd{}{r}\left(r\pd{\phi^i}{R}\right) + \frac{\veps^3\hat{\sigma}^i}{ r}\pd{}{R}\left(r\pd{\phi^i}{r}\right) + \veps^2\hat{\sigma}^i\spd{\phi}{R}} \hspace{5cm}&\label{eqn:nondimLaplacea12} \\
&=- \veps^2 F^i(r,R-\theta/2\pi), \qquad\qquad\non             \quad i \in \{ a_1, a_2\}
	,\end{align}
with the solution periodic in $R$ and $\theta$ with period $1$ and $2 \pi$ respectively. Taking the unit cell to be $-\delta^+<R-R_0-\theta/2 \pi<1-\delta^+$, the boundary conditions  \eref{eq:nondimBndconds1}-\eref{eq:nondimBndconds4} become 
\begin{align}
\left.\phi^+\right|_{R = R_0 + \delta^+ + \pdhfrac{\theta}{2\pi}} &= 
\left.	\phi^{a_1} \right|_{R = R_0 + \delta^+ + \pdhfrac{\theta}{2\pi}}
	\label{eq:conditions1a}\\
\left.\phi^+\right|_{R = R_0  - \delta^+ + \pdhfrac{\theta}{2\pi}} &= 
\left.	\phi^{a_2} \right|_{R = R_0 + 1 - \delta^+ + \pdhfrac{\theta}{2\pi}},		
	\label{eq:conditions1b}\\
	\phi^- &= 
	\begin{cases}
	\phi^{a_1} & \text{on} \ R = R_0 + \sep - \delta^- + \pdhfrac{\theta}{2\pi},\\
	\phi^{a_2} & \text{on} \ R = R_0 + \sep + \delta^- + \pdhfrac{\theta}{2\pi},		
	\end{cases}
	\label{eq:conditions2}\\
 \left.\sigma^+	\vec{n} \cdot \del \phi^+ \right|_{R = R_0 + \delta^+ + \pdhfrac{\theta}{2\pi}}&= 
	\left.	\varepsilon^4 \hat{\sigma}^{a_1}\vec{n} \cdot \del\phi^{a_1}\right|_{R = R_0 + \delta^+ + \pdhfrac{\theta}{2\pi}}
	\label{eq:conditions3a}\\
        \left.\sigma^+	\vec{n} \cdot \del \phi^+ \right|_{R = R_0
          - \delta^+ + \pdhfrac{\theta}{2\pi}}&= 
	\left.	\varepsilon^4 \hat{\sigma}^{a_2}\vec{n} \cdot \del\phi^{a_2} \right|_{R = R_0 + 1 - \delta^+ + \pdhfrac{\theta}{2\pi}},
	\label{eq:conditions3b}\\
\sigma^-	\vec{n} \cdot \del\phi^- &= 
	\begin{cases}
		\varepsilon^4 \hat{\sigma}^{a_1}\vec{n} \cdot \del\phi^{a_1} & \text{on} \  R = R_0 + \sep - \delta^- + \pdhfrac{\theta}{2\pi},\\
		\varepsilon^4 \hat{\sigma}^{a_2}\vec{n} \cdot \del\phi^{a_2} & \text{on} \  R = R_0 + \sep + \delta^- + \pdhfrac{\theta}{2\pi},
	\end{cases}
	\label{eq:conditions4}
\end{align}
where $R_0$ is the fractional part of $r_0/\varepsilon$, and,  using  the expression derived in \eref{eqn:normal}, $\vec{n} \cdot \del\phi$ can be written in multiple-scales form as
\begin{align}
	\left(-\frac{\varepsilon}{2 \pi r^2}\pd{\phi}{\theta} + \pd{\phi}{r} + \frac{1}{\varepsilon}\pd{\phi}{R}\right)\left(1+\frac{\varepsilon^2}{4 \pi^2 r^2} \right)^{-1/2};\label{eq:normal}
\end{align}
the square-root factor in this expression will cancel in all the relevant boundary conditions.

As usual we expand the solution in powers of $\varepsilon$ in the form
\begin{equation}
\phi^i \sim \phi_0^i + \varepsilon \phi_1^i + \varepsilon^2 \phi_2^i + \ldots,
\end{equation}
and equate coefficients of powers of $\varepsilon$.
%
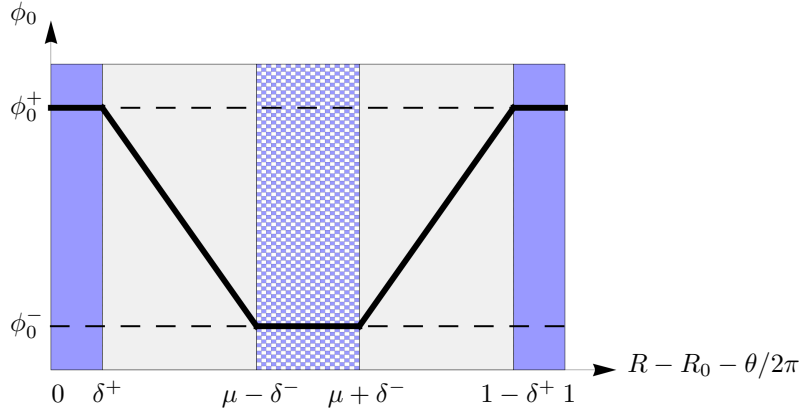
\begin{figure}
    \centering
    \begin{overpic}[trim=0 -3cm 0 0,width=0.6\textwidth]{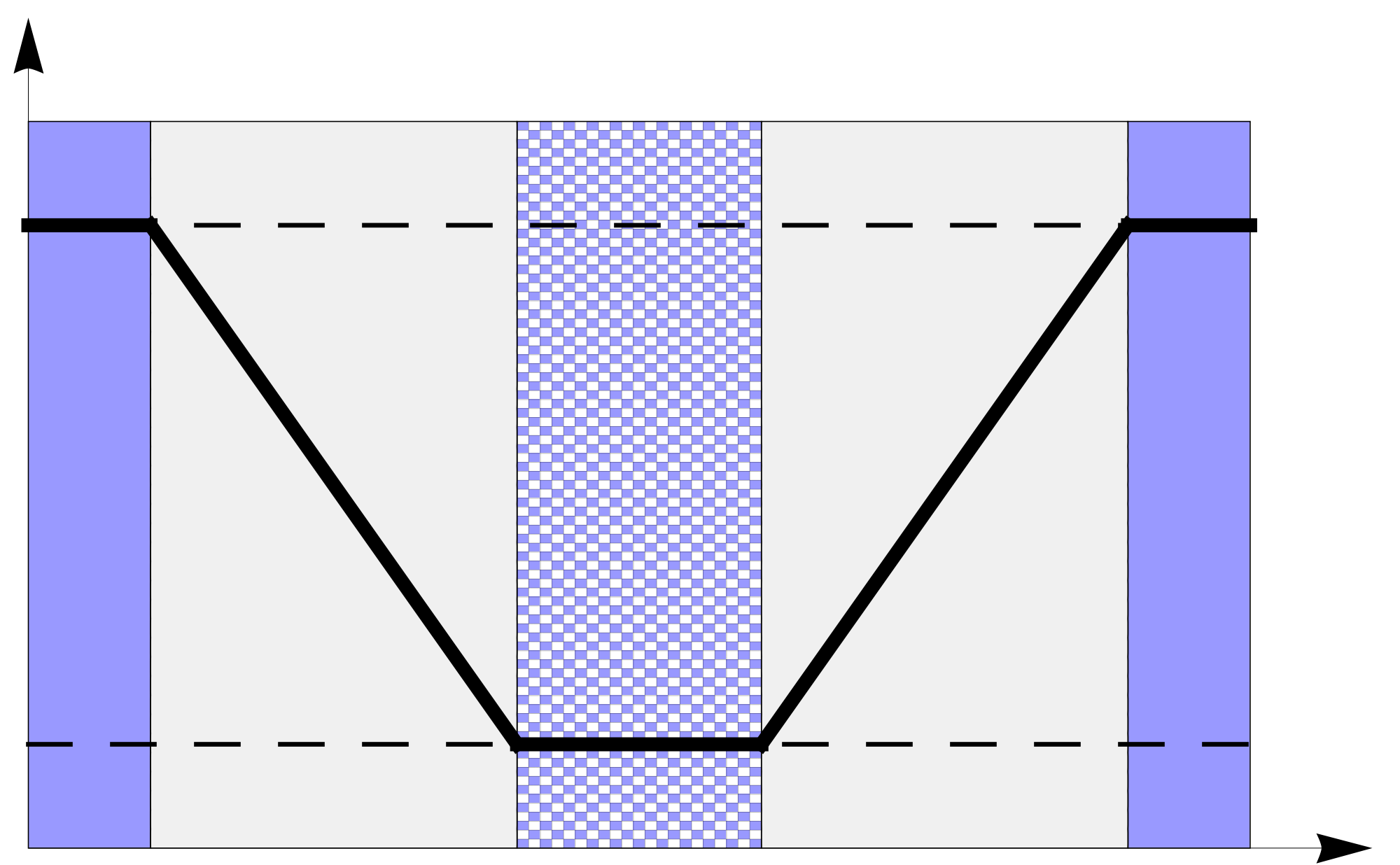}
      \put(100,5){$R-R_0-\theta/2\pi$}
      \put(2,0){$0$}
      \put(9,0){$\delta^+$}
      \put(31,0){\rotatebox{0}{$\sep-\delta^-$}}
      \put(49,0){$\sep+\delta^-$}
      \put(75,0){$1-\delta^+$}
      \put(89,0){$1$}
      
     \put(-5,65){$\phi_0$}
      \put(-5,49){$\phi_0^+$}
      \put(-5,12){$\phi_0^-$}
      \end{overpic}
    \caption{Representative leading order solution in the current collectors and active material regions, taken in the perpendicular direction. The current collectors are at constant (but distinct) potentials, while the potential varies linearly through the active material.}
    \label{fig:cell_soln}
\end{figure}
We begin with the equations in the positive current collector.
At leading order in \eref{eqn:nondimLaplacepm}, \eref{eq:conditions3a}-\eref{eq:conditions3b} we find
\begin{align}
	\spd{\phi_0^+}{R} &= 0,\\
	\pd{\phi_0^+}{R} &= 0 \ \text{on} \ R = R_0 \pm \delta^+ + \frac{\theta}{2\pi}.
\end{align}
Periodicity in $R$ implies that $\phi_0^+$ is independent of $R$, that is,
\begin{equation}
	\phi_0^+ = \phi_0^+(r, \theta).
	\label{eq:phi0p}
\end{equation}
At next order in \eref{eqn:nondimLaplacepm}, \eref{eq:conditions3a}-\eref{eq:conditions3b} we find
\begin{align}
 \mpd{\phi_0^+}{r}{R}+ \spd{\phi_1^+}{R} &= 0,\label{eq:order1rad}\\
\pd{\phi_0^+}{r} + \pd{\phi_1^+}{R} &= 0 \ \text{on} \ R = R_0 \pm \delta^+ + \frac{\theta}{2\pi}.\label{bc:order1rad}
\end{align}
Integrating \eref{eq:order1rad} and applying \eref{bc:order1rad} leads to
\begin{equation}
\phi_1^+ = -\pd{\phi_0^+}{r}\left(R - R_0 - \frac{\theta}{2\pi}\right) + \beta_1^+,
\label{eq:phi1p}
\end{equation}
where $\beta_1^+(r, \theta)$ is an arbitrary function of integration, and the terms $R_0+\theta/2\pi$ are included for convenience. At next order in \eref{eqn:nondimLaplacepm}, after simplifying using \eref{eq:phi1p}, we find
\begin{equation}
	\spd{\phi_2^+}{R} = -\frac{1}{r^2}\spd{\phi_0^+}{\theta} + \spd{\phi_0^+}{r}.
	\label{eq:phi2rad2}
\end{equation}
Integrating \eref{eq:phi2rad2} twice gives
\begin{equation}
    \phi_2^+ = \frac{1}{2}\left(-\frac{1}{r^2}\spd{\phi_0^+}{\theta} + \spd{\phi_0^+}{r}\right)\left(R - R_0 - \frac{\theta}{2 \pi}\right)^2 + \alpha_2^+\left(R - R_0 - \frac{\theta}{2 \pi}\right) + \beta_2^+,
    \label{eq:phi2sol}
\end{equation}
where $\alpha_2^+(r, \theta)$ and $\beta_2^+(r, \theta)$ are arbitrary functions of integration.
Equating coefficients at next order in \eref{eq:conditions3a}-\eref{eq:conditions3b} gives the boundary conditions 
\begin{align}
  -\frac{1}{2\pi r^2}\pd{\phi_0^+}{\theta} \mp \frac{1}{r^2} \spd{\phi_0^+}{\theta}\delta^+ + \pd{\beta_1^+}{r} + \alpha_2^+= 0.\label{eq:bc2order}
\end{align}
Subtracting one from the other and imposing periodicity of $\phi_0^+$ in $\theta$ shows that $\phi_0^+$ is independent of $\theta$, and that
 $\alpha_2^+ = -\pd{\beta_1^+}{r}$. 
Hence
\begin{equation}
	\phi_2^+ = \frac{1}{2}\left(\sdd{\phi_0^+}{r}\right)\left(R - R_0 - \frac{\theta}{2 \pi}\right)^2 - \pd{\beta_1^+}{r}\left(R - R_0 - \frac{\theta}{2 \pi}\right) + \beta_2^+.
	\label{eq:phi2}
\end{equation}
A similar calculation at next order gives
\begin{align}
 \label{eq:phi3}  \phi_3^+ &= -\frac{1}{6}\frac{\d^3 \phi_0^+}{\d r^3}\left(R - R_0 - \frac{\theta}{2 \pi}\right)^3  + \frac{1}{2} \spd{\beta_1^+}{r}\left(R - R_0 - \frac{\theta}{2 \pi}\right)^2 \\
  &\mbox{ } \qquad\non + \left(\frac{1}{4 \pi^2 r^2} \fdd{\phi_0^+}{r} - \pd{\beta_2^+}{r}\right)\left(R - R_0 - \frac{\theta}{2 \pi}\right) + \beta_3^+
   ,
\end{align}
together with the requirement that $\beta_1^+$ is independent of $\theta$.
 At fourth order in (\ref{eqn:nondimLaplacepm}), \eref{eq:conditions3a}-\eref{eq:conditions3b} we find 
 \[
 \frac{1}{r}\pd{}{r}\left(r\pd{\phi_2^+}{r}\right) + \frac{1}{r^2}\spd{\phi_2^+}{\theta} +  \frac{1}{r}\left(\pd{}{r}\left(r\pd{\phi_3^+}{R}\right) + \pd{}{R}\left(r\pd{\phi_3
 	^+}{r}\right)\right) + \spd{\phi_4^+}{R} =- \frac{1}{\sigma^+}F^+,
 \]
with  boundary conditions
\begin{align}
\left.\left(-\frac{1}{2 \pi r^2}\pd{\phi_2^+}{\theta} + \pd{\phi_3^+}{r} + \pd{\phi_4^+}{R}\right)\right|_{R = R_0 + \delta^+ + \frac{\theta}{2 \pi}} = \frac{\hat{\sigma}^{a_1}}{\sigma^+}\left.\pd{\phi_0^{a_1}}{R}\right|_{R = R_0 + \delta^+ + \frac{\theta}{2 \pi}},\label{eq:O4bc1}\\
\left.\left(-\frac{1}{2 \pi r^2}\pd{\phi_2^+}{\theta} + \pd{\phi_3^+}{r} + \pd{\phi_4^+}{R}\right)\right|_{R = R_0 - \delta^+ + \frac{\theta}{2 \pi}} = \frac{\hat{\sigma}^{a_2}}{\sigma^+}\left.\pd{\phi_0^{a_2}}{R}\right|_{R = R_0+1 - \delta^+ + \frac{\theta}{2 \pi}}\label{eq:O4bc2},
\end{align}
which finally brings in both the  potential in the active material and the source term.
 Substituting for $\phi_2^+$ and $\phi_3^+$ and integrating once gives
 \begin{align}
 \label{eq:phi4}    \pd{\phi_4^+}{R} &= \frac{1}{6}\frac{\d^4\phi_0^+}{\d r^4}\left(R-R_0-\frac{\theta}{2 \pi}\right)^3 - \frac{1}{2}\frac{\d^3\beta_1^+}{\d r^3}\left(R-R_0-\frac{\theta}{2 \pi}\right)^2 \\
    &\mbox{ }\qquad- \left(\frac{3}{4 \pi^2 r} \fdd{}{r}\left(\frac{1}{r} \fdd{\phi_0^+}{r} \right) + \frac{1}{r^2}\spd{\beta_2^+}{\theta} - \spd{\beta_2^+}{r}\right)\left(R-R_0-\frac{\theta}{2 \pi}\right)\non \\
    &\mbox{ }\qquad + \alpha_4^+ - \frac{1}{\sigma^+}\int_{R_0}^{R-\theta/2\pi} F^+(r,y)\, \d y.
   \nonumber
\end{align}
 To impose (\ref{eq:O4bc1})-(\ref{eq:O4bc2})  we need the leading-order potential in the active material. At leading order in (\ref{eqn:nondimLaplacea12}) we find
 \begin{align}
\hat{\sigma}^{a_k}\spd{\phi_0^{a_k}}{R} &=-  F^{a_k}(r,R-\theta/2\pi), \qquad k = 1,2.
\end{align}
 so that
 \begin{align}
   \phi_0^{a_1} &=  \alpha_0^{a_1} R +\beta_0^{a_1}
   - \frac{1}{\hat{\sigma}^{a_1}}\int^{R-\theta/2\pi}_{R_0+\delta^+} (R-\theta/2\pi-y)F^{a_1}(r,y)\, \d y, \\
   \phi_0^{a_2} &=  \alpha_0^{a_2} R +\beta_0^{a_2}
   -  \frac{1}{\hat{\sigma}^{a_2}}\int^{R-\theta/2\pi}_{R_0+\sep+\delta^- } (R-\theta/2\pi-y)F^{a_2}(r,y)\, \d y.
\end{align}
 At leading order the continuity conditions  (\ref{eq:conditions1a})-(\ref{eq:conditions2}) give
\[
  \alpha_0^{a_1}(R_0+\delta^+ + \theta/2\pi)+\beta_0^{a_1}
     = \phi_0^+ = 
     \alpha_0^{a_2}(R_0+1-\delta^+ + \theta/2\pi)+ \beta_0^{a_2}- \atwo\hat{F}^{a_2}/\hat{\sigma}^{a_2},
     \]
     \[
       \alpha_0^{a_1}(R_0+\mu - \delta^- + \theta/2\pi)+\beta_0^{a_1} - \aone\hat{F}^{a_1}/\hat{\sigma}^{a_1}
   = \phi^-_0 = 
  \alpha_0^{a_2}(R_0+\mu+\delta^- + \theta/2\pi)+   \beta_0^{a_2},
\]
where
\beqa
\hat{F}^{a_1} & = & \frac{1}{\aone}\int^{R_0+\mu - \delta^- }_{R_0+\delta^+} (R_0+\mu - \delta^- -y)F^{a_1}(r,y)\, \d y,\\
\hat{F}^{a_2} & = & \frac{1}{\atwo}\int^{R_0+1-\delta^+ }_{R_0+\sep+\delta^- } (R_0+1-\delta^+ -y)F^{a_2}(r,y)\, \d y,
\eeqa
so that
\begin{align}
\label{eq:phia1}	\phi_0^{a_1} &= \left(\phi_0^- - \phi_0^+ + \frac{\aone\hat{F}^{a_1}}{\hat{\sigma}^{a_1}}\right)  \frac{\left( R - R_0 - \delta^+ - \frac{\theta}{2 \pi} \right)}{\sep-\delta^+- \delta^-} \\
        & \hspace{2cm}\mbox{ }\non+ \phi_0^+  - \frac{1}{\hat{\sigma}^{a_1}}\int^{R-\theta/2\pi}_{R_0+\delta^+} (R-\theta/2\pi-y)F^{a_1}(r,y)\, \d y,\\
	\phi_0^{a_2} &= \left(\phi_0^- - \phi_0^+-\frac{\atwo\hat{F}^{a_2}}{\hat{\sigma}^{a_2}}\right)  \frac{\left( R_0 + 1 - \delta^+ + \frac{\theta}{2 \pi} - R\right)}{1-\sep - \delta^+- \delta^-} + \phi_0^+ \label{eq:phia2}\\
        & \hspace{2cm}\mbox{ }+ \frac{\atwo\hat{F}^{a_2}}{\hat{\sigma}^{a_2}}
      -  \frac{1}{\hat{\sigma}^{a_2}}\int^{R-\theta/2\pi}_{R_0+\sep+\delta^- } (R-\theta/2\pi-y)F^{a_2}(r,y)\, \d y. \non
\end{align}
%
Subtracting \eref{eq:O4bc2} from \eref{eq:O4bc1} and using \eref{eq:phia1}-\eref{eq:phia2} and  \eref{eq:phi2}-\eref{eq:phi4} gives
\begin{align}
 \label{eq:cond1} \lefteqn{-\bar{F}^+ -\frac{\delta^+\sigma^+}{2 \pi^2} \frac{1}{r} \fdd{}{r}\left( \frac{1}{r} \fdd{\phi_0^+}{r}\right) - \frac{2 \delta^+\sigma^+}{r^2} \spd{\beta_2^+}{\theta} }\hspace{2cm}&\\
  &=
 \frac{\hat{\sigma}^{a_1} (\phi_0^- - \phi_0^+) + \aone\hat{F}^{a_1}}{\sep - \delta^+-\delta^-} +  \frac{\hat{\sigma}^{a_2} (\phi_0^- - \phi_0^+) - \atwo\hat{F}^{a_2}}{1-\sep - \delta^+-\delta^-}  + \bar{F}^{a_2},\non
\end{align}
where
\beq
\bar{F}^+ = \int_{R_0-\delta^+}^{R_0+\delta^- } F^+(r,y)\, \d y, \qquad \bar{F}^{a_2} = \int_{R_0 + \sep +\delta^- }^{R_0+1 - \delta^+ } F^{a_2}(r,y)\, \d y.
\eeq
Since \eref{eq:cond1} implies $\spd{\beta_2^+}{\theta}$ is a function of $r$ only,  $\beta_2^+$ must in fact be independent of $\theta$.
Thus
\begin{equation}
  -\frac{\delta^+\sigma^+}{2 \pi^2} \frac{1}{r} \fdd{}{r}\left( \frac{1}{r} \fdd{\phi_0^+}{r}\right)
=
\frac{\hat{\sigma}^{a_1} (\phi_0^- - \phi_0^+) }{\aone} +  \frac{\hat{\sigma}^{a_2} (\phi_0^- - \phi_0^+) }{\atwo}+ \hat{F}^{a_1}- \hat{F}^{a_2} + \bar{F}^+ + \bar{F}^{a_2}.\label{eq:positive}
\end{equation}
A similar analysis for the negative current collector gives
\begin{equation}
  -\frac{\delta^-\sigma^-}{2 \pi^2} \frac{1}{r} \fdd{}{r}\left( \frac{1}{r} \fdd{\phi_0^-}{r}\right)
  = \frac{\hat{\sigma}^{a_2} (\phi_0^+ - \phi_0^-) }{\atwo} +  \frac{\hat{\sigma}^{a_1} (\phi_0^+ - \phi_0^-) }{\aone}+ \hat{F}^{a_2}- \hat{F}^{a_1} + \bar{F}^-+ \bar{F}^{a_1},\label{eq:negative}
\end{equation}
where
\beq
\bar{F}^- = \int_{R_0+\sep-\delta^-}^{R_0+\sep+\delta^- } F^-(r,y)\, \d y,
\qquad \bar{F}^{a_1} = \int_{R_0+\delta^+ }^{R_0+\sep-\delta^- } F^{a_1}(r,y)\, \d y.
\eeq
By adding and subtracting appropriate multiples of (\ref{eq:positive})-(\ref{eq:negative}) we may alternatively write the following system of equations for the leading-order potentials in the positive and negative current collectors:
\begin{align}
 \label{eq:eqn1beta}\lefteqn{ \frac{1}{2 \pi^2}\frac{1}{r} \fdd{}{r}\left( \frac{1}{r} \fdd{}{r} \left( \phi_0^- - \phi_0^+ \right)\right) =\left(\frac{1}{\delta^+\sigma^+}  + \frac{1}{\delta^-\sigma^-}\right)\left(\frac{\hat{\sigma}^{a_1}}{\aone}  + \frac{\hat{\sigma}^{a_2}}{\atwo}\right) (\phi_0^- - \phi_0^+)}\qquad\qquad&\\
  &\qquad \qquad\mbox{ }+
  \frac{\hat{F}^{a_1}-\bar{F}^{a_1}  }{ \delta^- \sigma^-}
  -\frac{\hat{F}^{a_2}-\bar{F}^{a_2}  }{ \delta^+ \sigma^+}
  -\frac{ \bar{F}^- + \hat{F}^{a_2}}{ \delta^- \sigma^-}
  + \frac{ \bar{F}^+ + \hat{F}^{a_1}}{ \delta^+ \sigma^+}  \nonumber
  ,
\end{align}
\begin{align}
\frac{1}{2 \pi^2}	\frac{1}{r} \fdd{}{r}\left( \frac{1}{r} \fdd{}{r} \left(\delta^+\sigma^+ \phi_0^+ + \delta^-\sigma^-\phi_0^- \right)\right) &= -\bar{F},\label{eq:eqn2beta}
\end{align}
where
\[ \bar{F} = \bar{F}^++ \bar{F}^-+ \bar{F}^{a_1}+ \bar{F}^{a_2} = \int_0^1 F(r,y)\, \d y.\]


\subsubsection{Ohmic heating}
As noted in Section~\ref{sec:Ohmic}, we use the solution of the electrical problem to determine the source term in the thermal problem given by Ohmic heating. The leading order term in the current collectors is from the azimuthal current, so that
\beqas
 \sigma^{\pm} | \del \phi|^2 & \sim &
\frac{\veps^2\sigma^{\pm}}{r^2} \left( \pd{\phi_1^{\pm}}{\theta}
\right)^2 =\frac{\veps^2\sigma^{\pm}}{4 \pi^2 r^2} \left( \fdd{\phi_0^{\pm}}{r}
\right)^2 ,
\eeqas
while in the active material the Ohmic heating is
\beqas
 \veps^4 \hat{\sigma}^{a_k} | \del \phi|^2 &\sim& \veps^2 \hat{\sigma}^{a_k}  \left(\pd{\phi_0^{a_k}}{R}\right)^2 \sim\veps^2 \hat{\sigma}^{a_k} \frac{(\phi_0^--\phi_0^+)^2}{\ell_k^2}.
\eeqas

\subsection{Reasonably conductive current collectors}

\label{sec:resonably conductive_main}
The third distinguished limit is one in which the electrical conductivity of the current collector materials is reasonably large compared to that of the active material, specifically 
$\sigma^{a_k} = \varepsilon^2 \bar{\sigma}^{a_k}$, where $\bar{\sigma}^{a_k}=O(1)$ as $\veps\ra 0$, for $k=1,2$.

The asymptotic analysis of this case is very similar to the very conductive case of Section \ref{sec:very conductive}, and we give details in the supplementary material. The main difference is that the behaviour in the active region comes into the analysis of the current collectors at second order, even though we still have to continue to fourth order in the current collectors to obtain the final solvability condition. Although this makes the algebra more complicated, it does not significantly alter the process.

The results are summarised, along with the other limits, in the following section.
\section{Implications for practical methods}
\label{sec:review}
\subsection{Review of distinguished limits}


We have identified three distinct distinguished limits for a spirally wound cell in which the radius of the cell is much greater than the thickness of a layer, and the conductivity of the current collectors is much greater than that of the active material. Here we summarise these limits in dimensional form, translating back to the physical equations (\ref{eq:potential})-(\ref{eq:Thermal}) from the general problem (\ref{eqn:problem}), and interpret the resulting models.
We write the leading-order problem in each case, dropping the subscript $0$ for clarity.


\subsubsection{Electrical problem}~\\
\noindent
    {\bf (i) Poorly conductive current collectors.} With $\sigma^{+}\sim\sigma^{-}\sim \sigma^{a_1} \sim \sigma^{a_2}$, $h\ll L$ we find
\begin{equation}
      \frac{\sigma_N}{r}\pd{}{r}\left( r \pd{\phi}{r}
    \right) +
    \frac{\sigma_T}{r^2} \PDD{\phi}{\theta} = 0,
         \label{eq:summary_poor}
\end{equation}
where
\[\sigma_N  = 
    \frac{1}{2\delta^+/\sigma^+ + 2 \delta^-/\sigma^- +\aone/\sigma^{a_1}+\atwo/\sigma^{a_2}}
  ,\quad
    \sigma_T  =  
    2\delta^+\sigma^{+} + 2\delta^-\sigma^{-} + \aone\sigma^{a_1} + \atwo\sigma^{a_2}.
    \]
In this case the spiral structure is irrelevant---only the local anisotropy of the material matters, not the global connectivity. The radial conductivity essentially arises from conductors in series, and the azimuthal conductivity essentially arises from conductors in parallel.
The Ohmic heating in each region is
\beq
F^i  =  
\frac{\sigma_N^2}{\sigma^i} \left( \pd{\phi}{r}\right)^2+\frac{\sigma^i}{r^2} \left(\pd{\phi}{\theta}\right)^2 , \qquad i \in\{+,-,a_1,a_2\}.
\label{ohmicpoor}
\eeq

\noindent
    {\bf (ii) Reasonably conductive current collectors.} With 
  $L^2 \sigma^{a_1}\sim L^2 \sigma^{a_2} \sim h^2 \sigma^+ \sim h^2 \sigma^-$, $h\ll L$, we find
 \begin{equation}
\frac{1}{(\aone/\sigma^{a_1} + \atwo/\sigma^{a_2})}	\frac{1}{r}\fdd{}{r}\left(  r\fdd{\phi}{r} \right)+	\frac{h^2(2\delta^+\sigma^+ +2\delta^-\sigma^-)}{(2 \pi)^2r}\fdd{}{r}\left(  \frac{1}{ r}  \fdd{\phi}{r} \right)= 0.
	\label{eq:summary_reasonable}
\end{equation}
 In this case $\phi$ is independent of $\theta$, and there is an enhanced radial conductivity which arises from the spiral winding---current can move in the radial direction both through the active material (short distance, low conductivity) and by flowing along the current collectors (long distance, high conductivity), with the resistances of these two paths comparable.
 The conductivity through the active material
 again corresponds to conductors in series,
 while the conductivity of the additional path through the current collectors corresponds to conductors in parallel. The additional twist is the unusual combination of radial derivatives in this term; these arise because
 \[	\frac{h^2(2\delta^+\sigma^+ +2\delta^-\sigma^-)}{(2 \pi)^2}\frac{1}{r}\fdd{}{r}\left(  \frac{1}{ r}  \fdd{\phi}{r} \right) =  (2\delta^+\sigma^+ +2\delta^-\sigma^-)\sdd{\phi}{s}\]
where
\[s = \frac{1}{h}\int_{r_0}^r 2 \pi r' \, \d r' \]
is arc length along the spiral winding.

The Ohmic heating is
\beq
F^{\pm} = \frac{h^2\sigma^{\pm}}{4 \pi^2 r^2} \left( \fdd{\phi}{r}
\right)^2 ,\label{ohmicmedium} \qquad
F^{a_k} = 
\frac{1}{\sigma^{a_k} }\frac{1}{(\aone /\sigma^{a_1} + \atwo /\sigma^{a_2})^2}\left( \fdd{\phi}{r}
\right)^2, ~k =1,2,
\eeq
in the current collectors and active material respectively.

\noindent
    {\bf (iii) Very conductive current collectors.} With $L^4 \sigma^{a_1}\sim L^4 \sigma^{a_2} \sim h^4 \sigma^+ \sim h^4 \sigma^-$, $h\ll L$ we find
\begin{align}
  \frac{h^2\delta^+\sigma^+}{2 \pi^2} \frac{1}{r} \fdd{}{r}\left( \frac{1}{r} \fdd{\phi^+}{r}\right) +\frac{\sigma^{a_1} (\phi^- - \phi^+) }{\aone h^2} + \frac{\sigma^{a_2} (\phi^- - \phi^+) }{\atwo h^2}
&=
 0,\label{eq:summary_very_positive}\\
  \frac{h^2 \delta^-\sigma^-}{2 \pi^2} \frac{1}{r} \fdd{}{r}\left( \frac{1}{r} \fdd{\phi^-}{r}\right)+ \frac{\sigma^{a_2} (\phi^+ - \phi^-) }{\atwo h^2} + \frac{\sigma^{a_1} (\phi^+ - \phi^-) }{\aone h^2}
  & = 0.\label{eq:summary_very_negative}
\end{align}
In this case not only is the spiral winding important, but also the fact that it is double wound, i.e. there are two current collectors---the potential takes different values in the positive and negative current collectors, each of which is independent of $\theta$. The dominant path for current flow is along the current collectors, but the potential difference between them also drives a current through the active material.  

The Ohmic heating  is
\beq
F^{\pm} = \frac{h^2\sigma^{\pm}}{4 \pi^2 r^2} \left( \fdd{\phi^{\pm}}{r}
\right)^2 ,\label{ohmicvery}\qquad 
 F^{a_k}  =  \sigma^{a_k} \frac{(\phi^--\phi^+)^2}{\ell_k^2 h^2}, ~k=1,2,
\eeq
in the current collectors and active material respectively.

\subsubsection{Thermal problem}~

\noindent
 {\bf (i) Poorly thermally conductive current collectors.} With $k^{+}\sim k^{-}\sim k^{a_1} \sim k^{a_2}$; $h\ll L$.
\begin{equation}
      \frac{k_N}{r}\pd{}{r}\left( r \pd{T}{r}
    \right) +
    \frac{k_T}{r^2} \PDD{T}{\theta} = \bar{F},
         \label{eq:summary_poor_thermal}
\end{equation}
where
\[
k_N  = 
    \frac{1}{2\delta^+/k^+ + 2 \delta^-/k^- +\aone/k^{a_1}+\atwo/k^{a_2}}
  ,\quad
    k_T  =  
    2\delta^+k^{+} + 2\delta^-k^{-} + \aone k^{a_1} + \atwo k^{a_2},
\]
and
\[    \bar{F}  =   \int_0^1 F(r,y)\d y,
\]
where $F$ is given by the appropriate electrical problem, i.e. (\ref{ohmicpoor}), (\ref{ohmicmedium}) or (\ref{ohmicvery}).
Since in each case $F$ is independent of $R-\theta/2\pi$ in each material,
we have
\begin{align}
\bar{F} & =  \sigma_N \left( \pd{\phi}{r}\right)^2+\sigma_T \left(\frac{1}{r}\pd{\phi}{\theta}\right)^2,\label{F1}\\
\bar{F} & =\left(\frac{h^2(2\delta^+\sigma^{+}+2\delta^-\sigma^-)}{4 \pi^2 r^2} + \frac{1}{(\aone /\sigma^{a_1} + \atwo /\sigma^{a_2})}\right)\left( \fdd{\phi}{r}
\right)^2,\label{F2}\\
\bar{F} & =\frac{2\delta^+\sigma^{+}h^2}{4 \pi^2 r^2}\left( \fdd{\phi^+}{r}
\right)^2 +\frac{2\delta^-\sigma^{-}h^2}{4 \pi^2 r^2}\left( \fdd{\phi^-}{r}
\right)^2 + \left(\frac{\sigma^{a_1}}{\aone} + \frac{\sigma^{a_2}}{\atwo}\right)\frac{(\phi^--\phi^+)^2}{h^2},\label{F3}
\end{align}
in electrical cases (i), (ii) and (iii) respectively.

 \noindent
 {\bf (ii)  Reasonably thermally conductive current collectors.} With 
  $L^2 k^{a_1}\sim L^2 k^{a_2} \sim h^2 k^+ \sim h^2 k^-$, $h\ll L$, we find
 \begin{equation}
\frac{1}{(\aone/k^{a_1} + \atwo/k^{a_2})}	\frac{1}{r}\fdd{}{r}\left(  r\fdd{T}{r} \right)+	\frac{h^2(2\delta^+k^+ +2\delta^-k^-)}{(2 \pi)^2r}\fdd{}{r}\left(  \frac{1}{ r}  \fdd{T}{r} \right)
	= \bar{F},
	\label{eq:summary_reasonable_thermal}
 \end{equation}
 with $\bar{F}$ again given by one of (\ref{F1})-(\ref{F3}).

 \noindent
     {\bf (iii) Very thermally conductive current collectors.} With $L^4 k^{a_1}\sim L^4 k^{a_2} \sim h^4 k^+ \sim h^4 k^-$, $h\ll L$ we find
\begin{align}
  \frac{h^2\delta^+k^+}{2 \pi^2} \frac{1}{r} \fdd{}{r}\left( \frac{1}{r} \fdd{T^+}{r}\right) +\frac{k^{a_1} (T^- - T^+) }{\aone h^2} + \frac{k^{a_2} (T^- - T^+) }{\atwo h^2}
&=S_1 ,\label{eq:summary_very_positive_thermal}\\
  \frac{h^2 \delta^-k^-}{2 \pi^2} \frac{1}{r} \fdd{}{r}\left( \frac{1}{r} \fdd{T^-}{r}\right)+ \frac{k^{a_2} (T^+ - T^-) }{\atwo h^2} + \frac{k^{a_1} (T^+ - T^-) }{\aone h^2 }
  & = S_2 ,\label{eq:summary_very_negative_thermal}
\end{align}
where $S_1 =  \hat{F}^{a_1}- \hat{F}^{a_2} + \bar{F}^+ + \bar{F}^{a_2}$, $S_2 = \hat{F}^{a_2}- \hat{F}^{a_1} + \bar{F}^-+ \bar{F}^{a_1}$,  
\beqa
\hat{F}^{a_1} & = & \frac{1}{\aone}\int^{R_0+\mu - \delta^- }_{R_0+\delta^+} (R_0+\mu - \delta^- -y)F^{a_1}(r,y)\, \d y,\\
\hat{F}^{a_2} & = & \frac{1}{\atwo}\int^{R_0+1-\delta^+ }_{R_0+\sep+\delta^- } (R_0+1-\delta^+ -y)F^{a_2}(r,y)\, \d y,
\eeqa
and $\bar{F}^i$ is the integral of $F$ over the corresponding region, with $F$ as given by the appropriate electrical problem, i.e. (\ref{ohmicpoor}), (\ref{ohmicmedium}) or (\ref{ohmicvery}). 
Since in each case $F$ is independent of $R-\theta/2\pi$ in each material,
we have $\hat{F}^{a_k}  = \bar{F}^{a_k}/2$, so that
\[
S_1 =   \frac{\bar{F}^{a_1}}{2}+ \frac{\bar{F}^{a_2}}{2} + \bar{F}^+,\qquad
S_2  =   \frac{\bar{F}^{a_1}}{2}+ \frac{\bar{F}^{a_2}}{2} + \bar{F}^-;
\]
with uniform heating half the heat generated in each active region travels to the positive current collector, and half to the negative current collector.
Explicitly, we have
\begin{align*}
  S_1 & =  \sigma_N^2\left(\frac{2 \delta^+}{\sigma^+} + \frac{\aone}{2 \sigma^{a_1}} + \frac{\atwo}{2 \sigma^{a_2}} \right) \left( \pd{\phi}{r}\right)^2+\frac{(4 \delta^+\sigma^+ + \aone \sigma^{a_1}+\atwo \sigma^{a_2})}{2r^2}  \left(\pd{\phi}{\theta}\right)^2\!\!,   \\
  S_2 & =   \sigma_N^2\left(\frac{2 \delta^-}{\sigma^-} + \frac{\aone}{2 \sigma^{a_1}} + \frac{\atwo}{2 \sigma^{a_2}} \right) \left( \pd{\phi}{r}\right)^2+\frac{(4 \delta^-\sigma^- + \aone \sigma^{a_1}+\atwo \sigma^{a_2})}{2r^2}  \left(\pd{\phi}{\theta}\right)^2\!\!,  
  \\[1mm]
  S_1 & =   \frac{2 \delta^+h^2\sigma^{+}}{4 \pi^2 r^2} \left( \fdd{\phi}{r}
\right)^2+ \frac{1}{2(\aone /\sigma^{a_1} + \atwo /\sigma^{a_2})}\left( \fdd{\phi}{r}
\right)^2 ,   \\
  S_2 & =  \frac{2 \delta^-h^2\sigma^{-}}{4 \pi^2 r^2} \left( \fdd{\phi}{r}
\right)^2 + \frac{1}{2(\aone /\sigma^{a_1} + \atwo /\sigma^{a_2})}\left( \fdd{\phi}{r}
\right)^2 ,
   \\[1mm]
  S_1 & =   \frac{2 \delta^+h^2\sigma^{+}}{4 \pi^2 r^2} \left( \fdd{\phi^+}{r}
\right)^2 + \frac{1}{2}\left(\frac{\sigma^{a_1}}{\aone} + \frac{\sigma^{a_2}}{\atwo}\right)\frac{(\phi^--\phi^+)^2}{h^2}, \\  
  S_2 & =  \frac{2 \delta^-h^2\sigma^{-}}{4 \pi^2 r^2} \left( \fdd{\phi^-}{r}
\right)^2  + \frac{1}{2}\left(\frac{\sigma^{a_1}}{\aone} + \frac{\sigma^{a_2}}{\atwo}\right)\frac{(\phi^--\phi^+)^2}{h^2} .
\end{align*}
in electrical cases (i), (ii) and (iii) respectively.



\subsection{Composite expansions that span distinguished limits}
It is interesting to compare the models derived here with one of the most common models exploited in the literature for battery thermal modelling (see \cite{Guo2014}, for example).
To illustrate the approach we start by locally aligning the coordinate system with the layers of the cell so that the  conductivity tensor can be written as
\begin{equation}
    {\bf K}_{\text{eff}} = \begin{pmatrix}
k_N & 0 \\
0 & k_T 
\end{pmatrix}
\label{eq:keff}
\end{equation}
where
$k_N$ and $k_T$
   are the conductivities normal (resistances in series) and tangential (resistances in parallel) to the layers respectively.
   To make the change to cylindrical coordinates  we need to locally rotate this tensor by a small angle, $\varepsilon/(2\pi r)$, representing the angle between the spiral and the circle, which transforms ${\bf K}_{\text{eff}}$ approximately to
   \begin{equation}
 {\bf k}_{\text{eff}} = \begin{pmatrix}
k_N+k_T(\varepsilon/(2\pi r))^2
& \varepsilon/(2\pi r)(k_N-k_T) \\
\varepsilon/(2\pi r)(k_N-k_T)
&k_T+ k_N(\varepsilon/(2\pi r))^2
\end{pmatrix}
\label{eq:keff_a}
\end{equation}
where we have ignored terms of $o(\varepsilon^2)$.
If $k_T \sim k_N$ then the effect of the rotation is minimal, and (\ref{eq:keff_a}) is the same as (\ref{eq:summary_poor_thermal}) at leading order.
However, if $k_N \sim \veps^2 k_T$ then (\ref{eq:keff_a}) forces the temperature to be independent of $\theta$ at leading order, and the leading-order radial conductivity  is the same as (\ref{eq:summary_reasonable_thermal}).
Thus (\ref{eq:keff_a}) provides a uniform composite approximation spanning
both the poorly conductive and reasonably conductive distinguished limits.
Note that when using this approach significant care must be taken to ensure the method remains  accurate when the conductivity is highly anisotropic, since
the dominant terms in the equation are the $\theta$-derivatives, and small errors in their approximation can completely swamp the radial behaviour.

A common approach to modelling the electrical behaviour of spirally wound cells is to ``unwind'' the cell, writing the potential as a function of arc length along the current collectors. The ``leakage current'' between current collectors then depends on the potential difference between them, but at different values of arc length (corresponding to one loop forward or one loop backward) \cite{Guo2014,Harb1999}. Translated into the radial coordinate this leads to the following forward-backward differential difference equations 
 \begin{align}
 \label{eq1comp}  \lefteqn{ \frac{\veps^2 \delta^+\sigma^+}{2 \pi^2} \frac{1}{r} \fdd{}{r}\left( \frac{1}{r} \fdd{\phi^+}{r}\right)+\frac{\sigma^{a_1} }{\aone \veps^2}  \left(\frac{(r+\veps/2)}{r}\phi^-(r+\veps/2) - \phi^+(r)\right)}\hspace{10cm}&& \\
   \non \mbox{ }+ \frac{\sigma^{a_2}  }{\atwo \veps^2}\left(\frac{(r-\veps/2)}{r}\phi^-(r-\veps/2) - \phi^+(r)\right)
&=
 0,\\
 \label{eq2comp}\lefteqn{  \frac{\veps^2 \delta^-\sigma^-}{2 \pi^2} \frac{1}{r} \fdd{}{r}\left( \frac{1}{r} \fdd{\phi^-}{r}\right)+ \frac{\sigma^{a_2}  }{\atwo \veps^2}\left(
\frac{(r+\veps/2)}{r} \phi^+(r+\veps/2) - \phi^-(r)\right) }\hspace{10cm} &&\\
 \non \mbox{ }+ \frac{\sigma^{a_1} }{\aone \veps^2} \left(\frac{(r-\veps/2)}{r}\phi^+(r-\veps/2) - \phi^-(r)\right)
  & = 0.
 \end{align}
 Note the factors $(r\pm\veps/2)/r$ which arise due to the increase in length of a circular arc with radius $r$---in (\ref{eq1comp}) for example these terms represent the length of the negative current collector per unit length of the positive current collector. These factors are missing in the model of \cite{Guo2014}, but since the parameters in \cite{Guo2014} fall towards the high conductivity regime the numerical consequence of the error is small.

 Equations (\ref{eq1comp})-(\ref{eq2comp}) give a model valid in both the reasonably conductive and very conductive regimes, which can be seen by Taylor expanding the nonlocal terms.
 However, a much simpler local composite model is 
 \begin{align}
   \label{eq:composite_positive}
   \lefteqn{\frac{1}{r}\fdd{}{r}\left(  r\left(\frac{1}{2(\aone/\sigma^{a_1} + \atwo/\sigma^{a_2})}+  \frac{2\delta^+\sigma^+h^2}{(2 \pi r)^2}\right)\fdd{\phi^+}{r} \right) }\hspace{10cm}&\\
    \mbox{ }+\frac{\sigma^{a_1} (\phi^- - \phi^+) }{\aone h^2} + \frac{\sigma^{a_2} (\phi^- - \phi^+) }{\atwo h^2}
&= 0,\non \\
 \label{eq:composite_negative}
 \lefteqn{	\frac{1}{r}\fdd{}{r}\left(  r\left( \frac{1}{2(\aone/\sigma^{a_1} + \atwo/\sigma^{a_2})}+  \frac{2 \delta^-\sigma^-h^2}{(2 \pi r)^2}
 \right)\fdd{\phi^-}{r} \right)}\hspace{10cm}&\\
   \mbox{ } + \frac{\sigma^{a_2} (\phi^+ - \phi^-) }{\atwo h^2} + \frac{\sigma^{a_1} (\phi^+ - \phi^-) }{\aone h^2}
  & = 0.\non
\end{align}
which also provides a uniform composite approximation spanning
both the reasonably  conductive and very conductive distinguished limits.
Care must be taken when the conductivity of the current collectors is towards the low end and the problem becomes ill-conditioned.

  
\subsection{Prismatic cells}

The observation that the very conductive limit is equivalent to a problem where the dominant conduction path is along the current collectors together with
a ``leakage current'' through the active material between current collectors  allows us to formulate homogenised macroscopic models for  other cell geometries without a detailed calculation.
In general we have 
\begin{align}
  \fdd{}{s}\left(2\delta^+\sigma^+ \fdd{\phi^+}{s}\right) +\frac{\sigma^{a_1} (\phi^- - \phi^+) }{\aone h^2} + \frac{\sigma^{a_2} (\phi^- - \phi^+) }{\atwo h^2}
&=
 0,\\
 \fdd{}{s}\left(2\delta^-\sigma^- \fdd{\phi^-}{s}\right) +
  \frac{\sigma^{a_2} (\phi^+ - \phi^-) }{\atwo h^2} + \frac{\sigma^{a_1} (\phi^+ - \phi^-) }{\aone h^2}
  & = 0,
\end{align}
where $s$ is arc length along the winding.

\begin{figure}[htbp]
\centering
\begin{overpic}[width=0.9\textwidth]{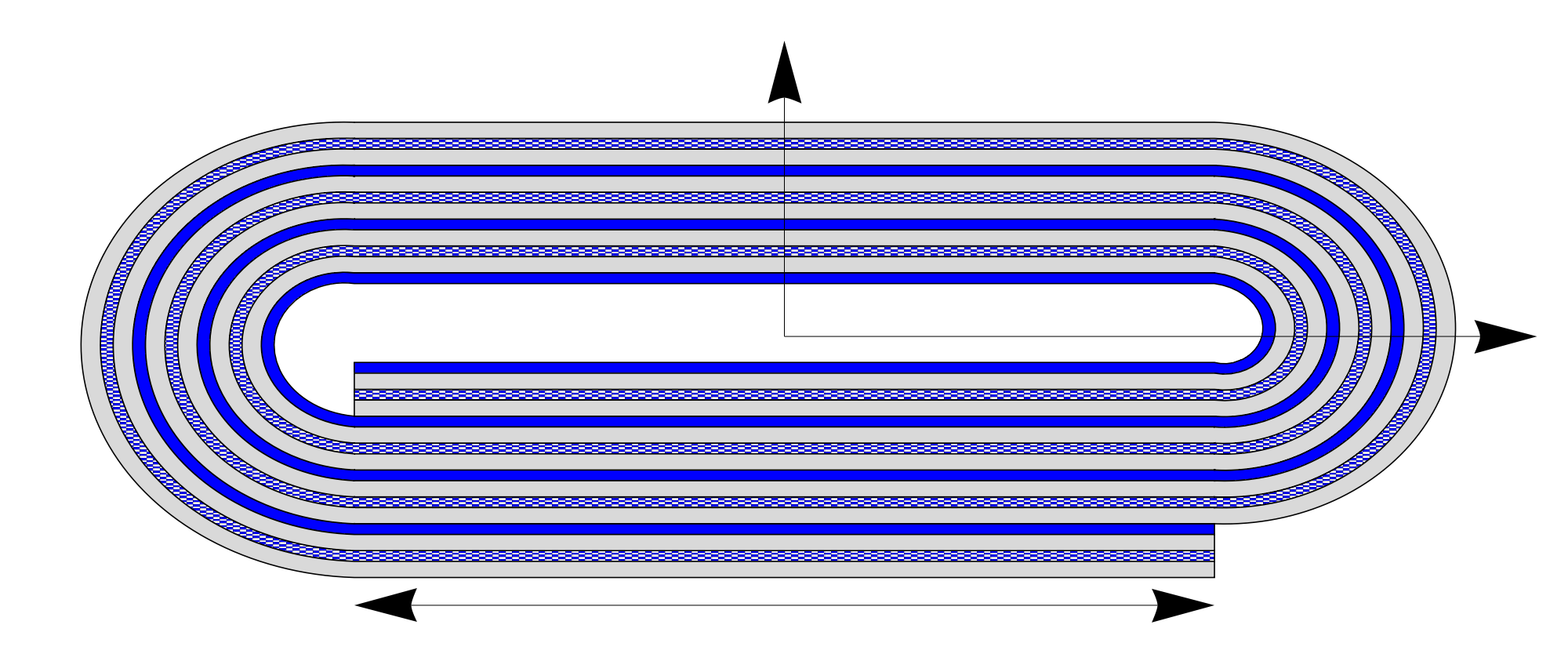}
  \put(96,18){$x$}
\put(47,38){$y$}
\put(50,1){$L$}
  \end{overpic}
    \caption{Representation of a prismatic pouch cell.}
    \label{fig:prismatic}
\end{figure}

Suppose we consider a long flat prismatic cell of width $L$ and thickness $H$, as illustrated in Figure \ref{fig:prismatic}.
Following the winding from position $(0,y)$ with $y>0$, arc length increases by $2L+2 \pi y$ (two traverses and two semi-circular end caps) by the time it returns to position $(0,y+h)$, where the vertical height has increased by the thickness $h$.
Thus the vertical coordinate $y$ and arc length $s$ are  related by
\[ \fdd{s}{y} = \frac{2L+2 \pi y}{h}, \quad y>0, \qquad
\fdd{s}{y} = -\frac{2L-2 \pi y}{h}, \quad y<0, 
\]
so that the homogenised model is
\begin{align}
 \frac{\delta^+\sigma^+h^2}{(L+\pi |y|)}\fdd{}{y}\left(  \frac{1}{2(L+ \pi |y|)} \fdd{\phi^+}{y}\right) +\frac{\sigma^{a_1} (\phi^- - \phi^+) }{\aone h^2} + \frac{\sigma^{a_2} (\phi^- - \phi^+) }{\atwo h^2}
&=
 0,\\
 \frac{ \delta^-\sigma^-h^2}{(L+\pi |y|)} \fdd{}{y}\left( \frac{1}{2(L+ \pi |y|)} \fdd{\phi^-}{y}\right)+
  \frac{\sigma^{a_2} (\phi^+ - \phi^-) }{\atwo h^2} + \frac{\sigma^{a_1} (\phi^+ - \phi^-) }{\aone h^2}
  & = 0,
\end{align}
with $\phi^\pm$ independent of the horizontal coordinate $x$.

\subsection{Boundary conditions}
Finally we need to consider how to apply boundary conditions to the homogenised problems that we have derived.  In the full electrical problem we might apply known potentials at certain points, representing the tabs on the current collectors, and require no current through any other external surfaces. For the thermal problem we might impose heat transfer conditions at the tabs and at the external surfaces (with differing coefficients) of the form $-k^i\nabla T\cdot\vec{n}=N(T-T_\infty)$ for a heat transfer coefficient $N$ and external temperature $T_\infty$.

For the poor conductor and the reasonable conductor limits the homogenised problem has a single potential/temperature. To find the homogenised boundary condition we assume that the condition varies on a length scale much larger than the microscale of the problem (this excludes situations such as applying the potential on the separate current collectors at two points very close to one another). If a Dirichlet condition is imposed then the homogenised potential/temperature has a Dirichlet condition, and similarly for a Neumann (no flux) condition. If a heat transfer condition is imposed the homogenised condition is found by replacing the thermal conductivity with the effective conductivity $k_{\text{eff}}$. 

For the very conductive case, Dirichlet conditions can be imposed on either current collector, as can any Neumann condition (note that unlike the poor and reasonably conductive cases the problem requires two conditions at every surface to be well-posed).  If a heat transfer condition is imposed then the conductivity is replaced by the effective conductivity due to that particular current collector, and the heat transfer coefficient is multiplied by the fraction of the surface belonging to that current collector.

In the reasonably conductive limit with a Dirichlet condition there is a boundary layer on the inner and outer surfaces that does not affect the leading order condition, but it does affect the numerical accuracy. We discuss an improved boundary condition in Section~\ref{sec:results}. 

\section{Comparison with COMSOL solutions}
\label{sec:results}
In this section we compare the solutions derived from the homogenised models for the spirally-wound cell for the electrical problem ($F=0$), with numerical solutions of the full problem solved directly on the original spiral geometry in COMSOL. To describe the various solutions we introduce the following  notation: $\phi_C$ is the potential obtained by solving the original problem in COMSOL;
$\phi_P$ is the potential obtained by solving (\ref{eq:summary_poor})
 for ``poorly conductive current collectors''; 
   $\phi_R$ is the potential obtained by solving (\ref{eq:summary_reasonable})  for ``reasonably conductive current collectors''; and 
 $\phi_V$ is the potential obtained by solving the (\ref{eq:summary_very_positive})-(\ref{eq:summary_very_negative}) for  ``very conductive current collectors''.
 For poor and reasonable conductors the solution of the homogenised problem directly gives the leading-order potential, which is uniform on the microscale. However, for very conductive current collectors the leading order potential
 varies linearly in the active material between the two values $\phi^+$ and $\phi^-$ obtained from the macroscale problem---see Fig.~\ref{fig:cell_soln}.
 Furthermore we use: 
 $\phi_{PR}$ for the potential obtained by using  \eref{eq:keff_a} which gives
 a composite solution valid across the poor and reasonably conductive regimes; and $\phi_{RV}$ for the potential obtained by solving the problem \eref{eq:composite_positive}--\eref{eq:composite_positive} which gives a composite solution valid across the reasonably and very conductive regimes. 

 For our comparisons we use parameters   representative of a 18650-type cylindrical cell, similar to that studied in \cite{Guo2014}.
 We choose a symmetric microstructure with $\sep=1/2$, $\delta^+ = \delta^-=0.05$,  $\sigma^+ = \sigma^-= \sigma^\pm$, and $\sigma^{a_1} = \sigma^{a_2} = \sigma^a$, say.
 We take $L_0 = 0.25 L$, and $\veps = 0.0375$, corresponding to 20 windings. We choose  $\sigma^{a}/\sigma^+ = \num{2e-7}$,
 representative of typical battery materials \cite{Lee2013}.
 In addition, to explore the validity of all the models we have derived, we keep the geometry fixed and consider three other values of  the conductivity ratio, chosen to correspond roughly to each of our distinguished limits. Broadly these are $\sigma^{a}/\sigma^\pm = 1$, $\veps^2$ and $\veps^4$, but since the formally $O(1)$  geometrical factors $\delta$ and $1/(4 \pi^2)$ are fairly small we take instead $\sigma^{a}/\sigma^\pm = 0.1$, $0.01 \veps^2$ and $0.01 \veps^4$.

\begin{figure}
    \centering
    \includegraphics[width=\textwidth]{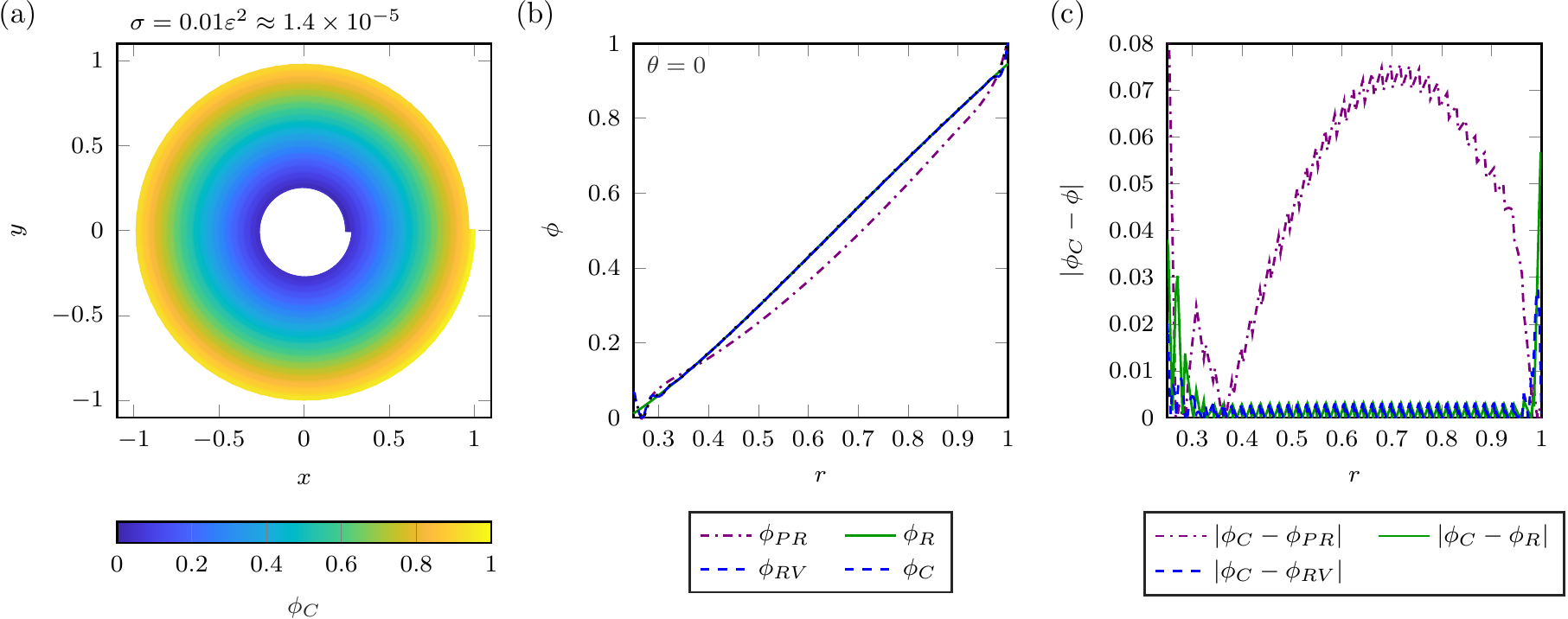}
    \caption{Numerical solutions of the case where $\sigma^a/\sigma^\pm= 0.01\varepsilon^2 \approx 1.4 \times 10^{-5}$, showing: (a) contour plots from COMSOL; (b) comparison of calculated homogenised potential with predictions at $\theta=0$ from COMSOL; (c) errors from comparisons.}
    \label{fig:compare2}
\end{figure}

 For most practical 18650 cells there would be several tabs, however here we choose a simple problem to allow easy comparison. We assume there is a single tab on the positive current collector at the outer radius and a single tab on the negative current collector at the inner radius. Hence we impose Dirichlet conditions $\phi^+=1$ at the outer radius (having scaled the potential with the  applied potential difference) and $\phi^- = 0$ at the inner radius. For the very high conductivity limit we need a second boundary condition: we  impose that no current leaves from the negative current collector at the external surface ($\partial\phi^-/\partial r = 0$ on the outer boundary) and no current leaves from the positive current collector at the centre of the jelly roll ($\partial\phi^+/\partial r=0$ on the inner boundary).
 For the reasonably conductive limit there is a boundary layer near each tab due to the asymmetry in the first and last winding (this can be seen in the COMSOL solution in Fig.~\ref{fig:compare2}(b)). 
 
 In Section SM2 we analyse these
 boundary layers and derive improved Robin boundary conditions which we use for $\phi_R$. In nondimensional variables these are
 \beq \fdd{\phi_R}{r}(r_0) = \frac{\alpha}{\veps} \phi_R(r_0), \qquad
 \fdd{\phi_R}{r}(1) = \frac{\beta}{\veps} (1-\phi_R(1)),\label{correctedbc}
 \eeq
 where $\alpha$ and $\beta$ depend on $\delta \ell h^2 \sigma^\pm/r_0^2\sigma^a$ and $ \delta \ell h^2 \sigma^\pm/\sigma^a$ respectively, and are given in the supplementary material. At leading order in $\veps$ these reduce to $\phi_R(1)=1$, $\phi_r(r_0)=0$ as described above, but we find that (\ref{correctedbc}) captures higher-order effects near the boundary leading to more accurate results. Note that these boundary layers would not be present if the tabs were somewhere in the middle rather than at the end of the current collectors, and that they are only important in the reasonably conductive regime. Similar boundary conditions are applied to the composite solution $\phi_{RV}$; these are given in the supplementary material.

Table~\ref{tab:errors} shows the maximum error in $r$ (at a fixed value $\theta = 0$) in the solutions of each of the homogenised limits relative to directly solving the problem on the spiral geometry in COMSOL for various values of the conductivity ratio $\sigma^a/\sigma^\pm$. This shows the general trend that as $\sigma^a/\sigma^\pm$ decreases the error in the poorly conductive models increases, and the error in the very conductive models decreases.
We see that the composite models are accurate over a wider range of values of $\sigma^a/\sigma^\pm$. In particular it is worth noting that the solution of the composite model spanning the reasonably conductive and very conductive limits, $\phi_{RV}$, performs remarkably well for all values $\sigma^a/\sigma^\pm \le 0.01 \varepsilon^2$, and always outperforms the solution of the reasonably conductive model, $\phi_R$.


\begin{table}
    \centering
    \begin{tabular}{c R R R R R}
        \toprule
 $\sigma^a/\sigma^\pm$ & $\scriptstyle  \max\limits_r|\phi_C - \phi_P|$ & $\scriptstyle  \max\limits_r|\phi_C - \phi_{PR}|$ & $\scriptstyle  \max\limits_r|\phi_C - \phi_R|$ & $\scriptstyle  \max\limits_r|\phi_C - \phi_{RV}|$ & $\scriptstyle  \max\limits_r|\phi_C - \phi_V|$  \EndTableHeader\\
        \midrule
        $0.1$ & 0.0426 & 0.0425 & 0.3054 & 0.3009 & 0.3018 \\
        $0.01 \varepsilon^2$ & 0.1666 & 0.0972 & 0.0568 & 0.0276 & 0.1181 \\
        $\num{2e-7}$ & 0.2075 & 0.2231 & 0.1908 & 0.0130 & 0.0132 \\
        $0.01 \varepsilon^4$ & 0.4395 & 0.4769 & 0.4446 & 0.0123 & 0.0123 \\
        \bottomrule
    \end{tabular}
    \caption{Maximum error of solutions of different homogenised limits compared to the COMSOL solution at $\theta=0$, relative to the applied potential difference, for various values of the conductivity ratio $\sigma^a/\sigma^\pm$. Numerically $0.01\varepsilon^2 \approx 1.4 \times 10^{-5}$ and $0.01\varepsilon^{4} \approx 1.9775 \times 10^{-8}$.
    }
    \label{tab:errors}
\end{table}

\begin{figure}
    \centering
    \includegraphics[width=\textwidth]{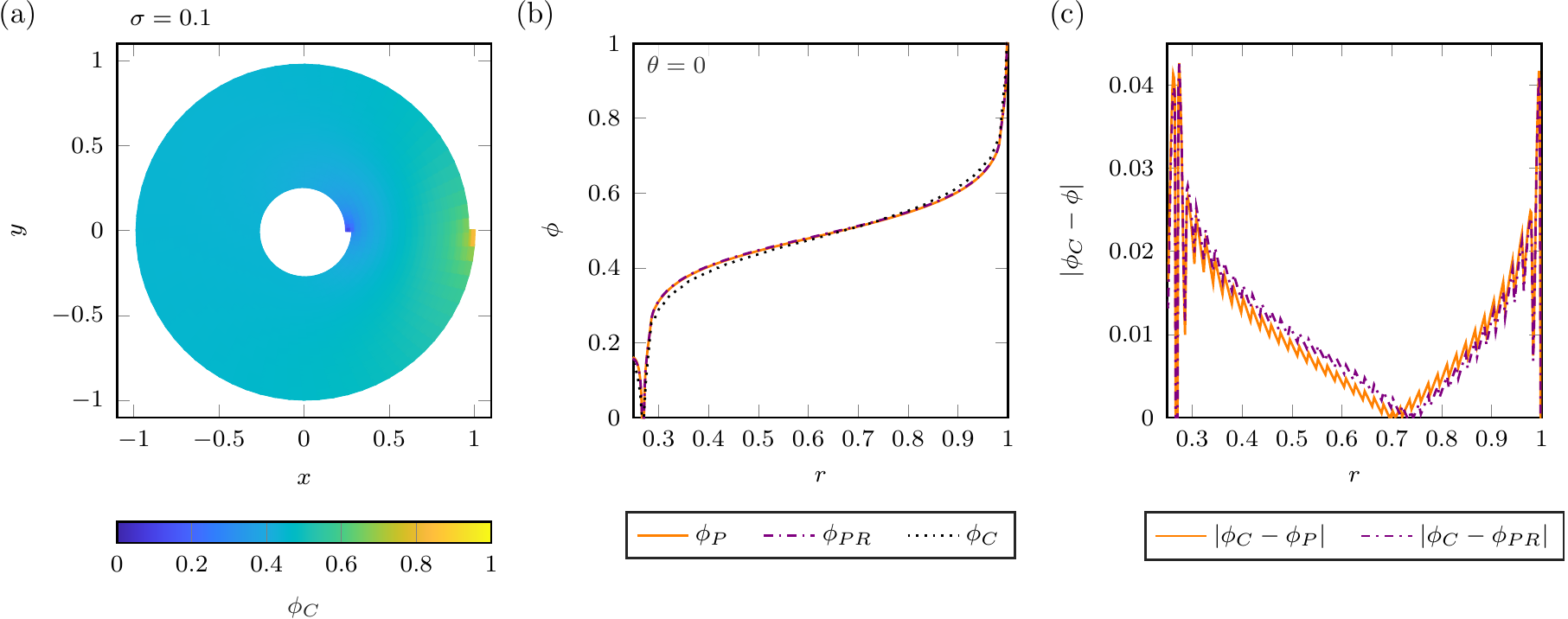}
    \caption{Numerical solutions of the case where $\sigma^a/\sigma^\pm=0.1$, showing: (a) contour plots from COMSOL; (b) comparison of calculated homogenised potential with predictions at $\theta=0$ from COMSOL; (c) errors from comparisons.}
    \label{fig:compare1}
\end{figure}

\begin{figure}
    \centering
    \includegraphics[width=\textwidth]{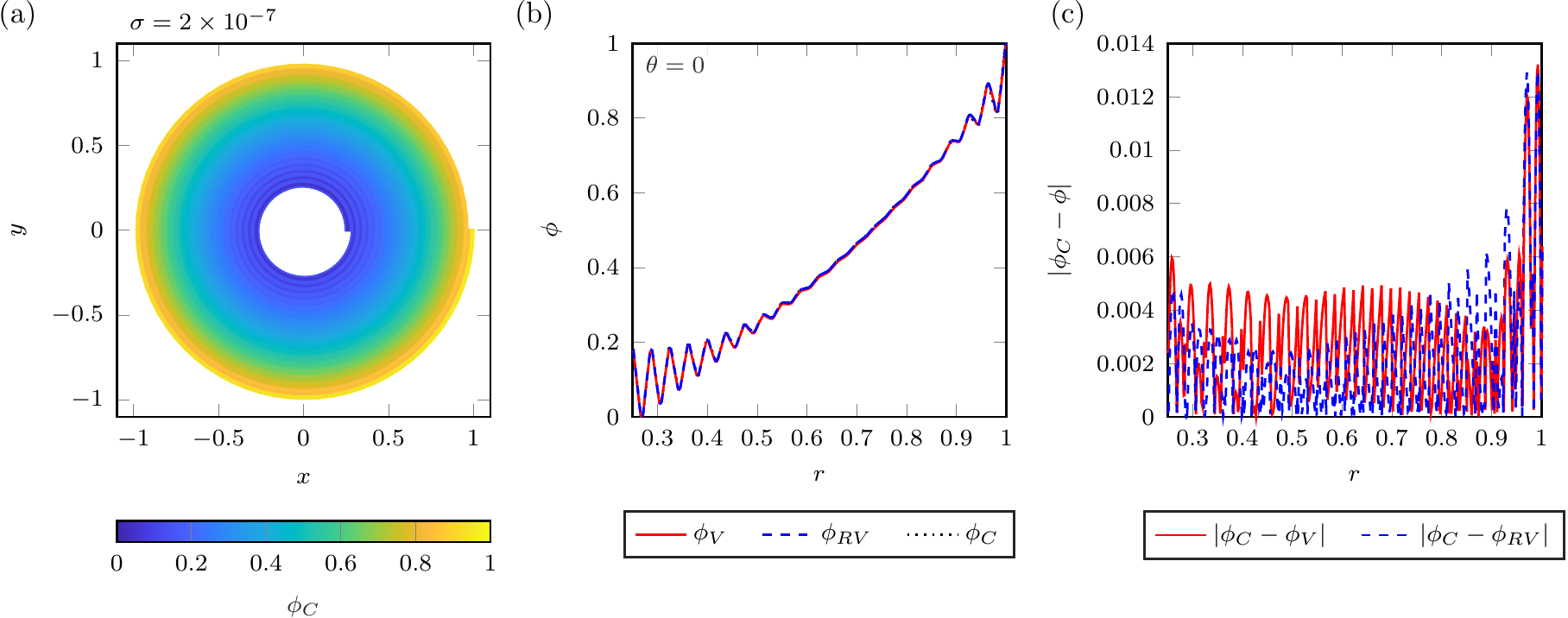}
        \caption{Numerical solutions of the case where $\sigma^a/\sigma^\pm = \num{2e-7}$, showing: (a) contour plots from COMSOL; (b) comparison of calculated homogenised potentials with predictions at $\theta=0$ from COMSOL; (c) errors from comparisons.}
    \label{fig:compare3}
\end{figure}

\begin{figure}
    \centering
    \includegraphics[width=\textwidth]{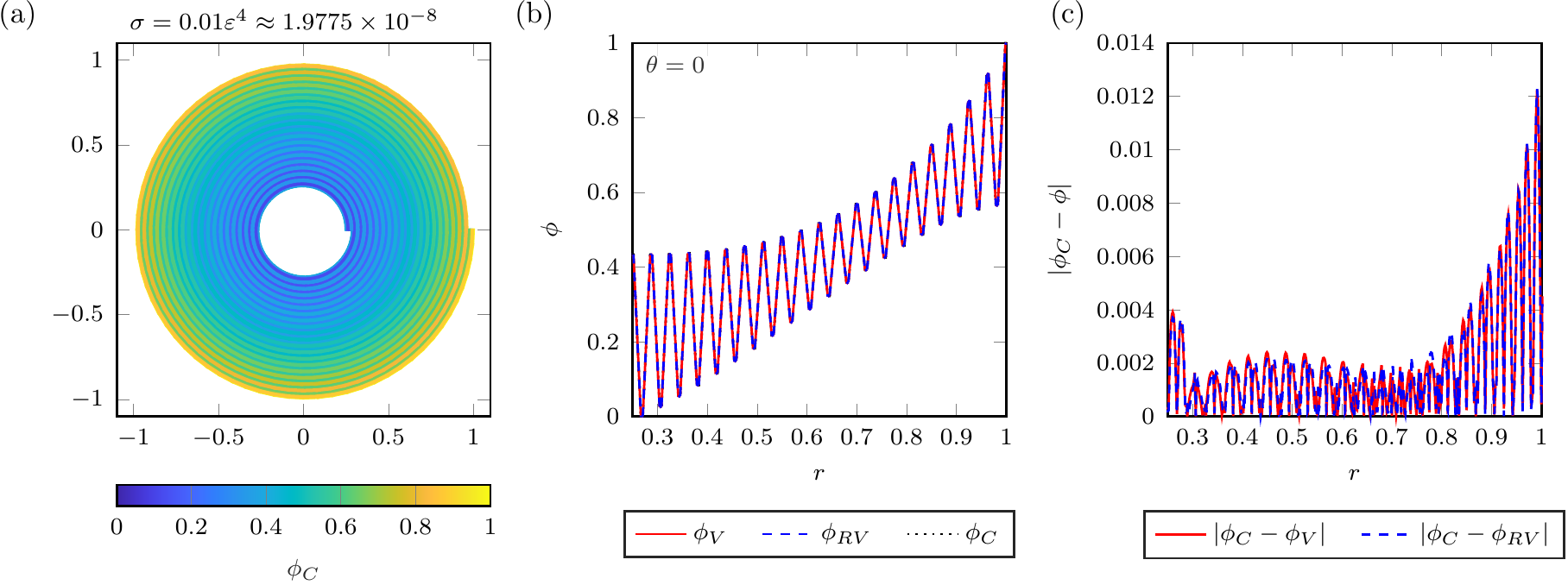}
    \caption{Numerical solutions of the case where $\sigma^a/\sigma^\pm= 0.01\varepsilon^4\approx\num{1.9775e-8}$, showing: (a) contour plots from COMSOL; (b) comparison of calculated homogenised potentials with predictions at $\theta=0$ from COMSOL; (c) errors from comparisons.}
    \label{fig:compare4}
\end{figure}

Figures~\ref{fig:compare1}--\ref{fig:compare4} provide a comparison of the numerical solutions of the full problem in COMSOL and the homogenised problems for various values of $\sigma^a/\sigma^\pm$, showing: (a) a contour plot of the potential from COMSOL; (b) the calculated potentials at $\theta=0$; and (c) the error in the homogenised solutions, also at $\theta=0$. 

As shown in Figure~\ref{fig:compare1}, the traditional homogenisation method accurately captures the potential when $\sigma^a/\sigma^\pm =0.1$.
Figure~\ref{fig:compare2} shows a comparison for $\sigma^a/\sigma^\pm = 0.01\varepsilon^2 \approx 1.4 \times 10^{-5}$, which roughly balances the terms in the reasonably conductive model. In this limit we expect there to be a single current collector potential (i.e. $\phi^+=\phi^-$)
which is true in the COMSOL solution apart from very close to the inner radius.
The small rapid fluctuations in potential are visible in the errors in 
Figure~\ref{fig:compare2}(c).

Figure~\ref{fig:compare3} shows the comparison for a conductivity ratio $\sigma^a/\sigma^\pm =  2 \times 10^{-7}$ typical of realistic battery materials \cite{Lee2013}. Here we observe the potential exhibits clear variations on the microscale close to the inner and outer boundaries (i.e. near the tabs), highlighting the importance of correctly accounting for the rapid variations in conductivity on the microscale. Both the very ($\phi_V$) and reasonably/very ($\phi_{RV}$) conductive models perform well, with maximum errors on the order of one percent.

Finally, Figure~\ref{fig:compare4} shows the results for $\sigma^a/\sigma^\pm=0.01\varepsilon^4$, which roughly balances the terms in the very conductive model. Here the potential varies on the microscale throughout the whole domain, rapidly jumping between its value on the negative and positive current collectors. In this limit the ``two-potential'' nature of the problem is clear, and it is obvious that using an effective conductivity tensor as derived in traditional homogenisation theory will be unable to produce the correct result. 


\section{Conclusions}
\label{sec:conclusion}
In this paper we have exploited the method of multiple-scales homogenisation to derive simplified models for the electrical and thermal behaviour of a spirally wound cell. The analysis identifies three distinguished limits for the conductivity
ratio relative to the lengthscale ratio. These  correspond to conventional homogenisation, in which there is an effective diffusion coefficient, a novel limit of radial diffusion, and a ``two-potential'' limit. These limits correspond to the physical cases of ``poorly conductive'' current collectors, ``reasonably conductive'' current collectors, and ``very conductive" current collectors respectively. The ``very conductive'' model is particularly novel, in that the  solution exhibits interesting variations on the microscale, and the homogenised problem involves not one but two functions of the macroscale. Each of the models accounts for the effect of the complicated spiral structure of the cell, but without the need to explicitly account for the structure in the numerical mesh. We also present two composite models that span the poorly/reasonably conductive cases and reasonably/very conductive cases, respectively. 

To allow for a clear exposition, the analysis assumed simple Ohmic behaviour for all cell components. However, the results obtained are general and could be readily extended to include a more realistic description of the electrochemistry (for example a locally one-dimensional computational model of behaviour between the current collectors), or indeed be applied to other physical problems. We have also indicated how the results may be extended in a straightforward manner to different cell geometries, such as prismatic cells.

A popular ad-hoc approach to modelling the electrical behaviour of spirally wound cells is to ``unroll'' the cell and solve for the potential along the length of the unwound current collector \cite{Harb1999,Guo2014}. The analysis here essentially validates such an approach in the high-conducting limit, but also demonstrates that
a much simpler problem, which we  posed along the radius instead of the arc length, can be solved instead. In contrast to the  pair of forward-backward differential-difference equations of \cite{Guo2014}, it is sufficient to solve simply a pair of ordinary differential equations.
The transformation to a radial coordinate may make coupling to other components such as temperature more straightforward, as well as facilitating a simple
reasonably/very conductive composite model. 

Although for our comparisons we have taken the tabs to be at the ends of the spirals, in reality  many cells have several tabs along the spirals. Conditions can be imposed to represent such  internal tabs, but if the lengthscale along which the tab connects to the current collector is much shorter than the cell dimension (i.e. much smaller than a turn of the spiral) there will be local effects on the solution. Such effects can be seen, for example, in the results of \cite{Lee2013}, and the model derived here may not be highly accurate in such cases. 

Comparisons between the solutions of the homogenised models with solutions obtained from directly solving the electrical problem on the spiral geometry in COMSOL were made for a range of parameter values. Through this, we demonstrated the ability of the homogenised models to accurately predict the potential, and quantified the errors made across the range of models. In particular it was shown that both the ``very conductive" model and ``reasonably/very conductive" composite model were able to predict the potential to within a few percent for realistic parameter values for a lithium-ion cell.
We also determined that the ``reasonably/very conductive" and the ``poor/reasonably conductive" composite models provide the greatest accuracy across the range of $\sigma$ values tested. 
To demonstrate the effectiveness of the modelling approach to  practical problems we examined  a typical 18650-type cell \cite{Lee2013,Guo2014}. We found that for 
the electrical problem, the values of conductivities indicate the ``reasonably/very conductive'' composite model is very suitable, while for the thermal problem the conductivities indicate the ``poor/reasonably conductive'' composite model should be used. These findings are in keeping with many other modelling approaches where a ``two-potential'' model is used for the electrical problem and an effective conductivity model is used for the thermal model.

\bibliographystyle{siamplain}
\bibliography{jellyrolls}

\begin{thebibliography}{10}

\bibitem{armand2008}
{\sc M.~Armand and J.~Tarascon}, {\em Building better batteries}, Nature, 451
  (2008), p.~652.

\bibitem{Chen2006}
{\sc S.-C. Chen, Y.-Y. Wang, and C.-C. Wan}, {\em Thermal analysis of spirally
  wound lithium batteries}, J. Electrochem. Soc., 153 (2006), pp.~A637--A648.

\bibitem{davit2013}
{\sc Y.~Davit, C.~Bell, H.~Byrne, L.~Chapman, L.~Kimpton, G.~Lang, K.~Leonard,
  J.~Oliver, N.~Pearson, R.~Shipley, S.~Waters, J.~Whiteley, B.~Wood, and
  M.~Quintard}, {\em Homogenization via formal multiscale asymptotics and
  volume averaging: How do the two techniques compare?}, Adv. Water Resour., 62
  (2013), pp.~178--206, \url{https://doi.org/10.1016/j.advwatres.2013.09.006}.

\bibitem{Doyle1993}
{\sc M.~Doyle, T.~Fuller, and J.~Newman}, {\em Modeling of galvanostatic charge
  and discharge of the lithium/polymer/insertion cell}, J. Electrochem. Soc.,
  140 (1993), pp.~1526--1533, \url{https://doi.org/10.1149/1.2221597}.

\bibitem{Duan2018}
{\sc X.~Duan, W.~Jiang, Y.~Zou, W.~Lei, and Z.~Ma}, {\em A coupled
  electrochemical--thermal--mechanical model for spiral-wound li-ion
  batteries}, J. Mater. Sci., 53 (2018), pp.~10987--11001.

\bibitem{Evans1989}
{\sc T.~Evans and R.~White}, {\em A thermal analysis of a spirally wound
  battery using a simple mathematical model}, J. Electrochem. Soc., 136 (1989),
  pp.~2145--2152.

\bibitem{Guo2014}
{\sc M.~Guo and R.~White}, {\em Mathematical model for a spirally-wound
  lithium-ion cell}, J. Power Sources, 250 (2014), pp.~220--235,
  \url{https://doi.org/10.1016/j.jpowsour.2013.11.023}.

\bibitem{Harb1999}
{\sc J.~Harb and R.~LaFollette}, {\em Mathematical model of the discharge
  behavior of a spirally wound {Lead‐Acid} cell}, J. Electrochem. Soc., 146
  (1999), pp.~809--818.

\bibitem{hunt2020}
{\sc M.~Hunt, F.~Brosa~Planella, F.~Theil, and W.~Widanage}, {\em Derivation of
  an effective thermal electrochemical model for porous electrode batteries
  using asymptotic homogenisation}, J. Eng. Math., 122 (2020), pp.~31--57.

\bibitem{kosch2018}
{\sc S.~Kosch, Y.~Zhao, J.~Sturm, J.~Schuster, G.~Mulder, E.~Ayerbe, and
  A.~Jossen}, {\em A computationally efficient multi-scale model for
  lithium-ion cells}, J. Electrochem. Soc., 165 (2018), pp.~A2374--A2388.

\bibitem{Lee2013}
{\sc K.-J. Lee, K.~Smith, A.~Pesaran, and G.-H. Kim}, {\em Three dimensional
  thermal-, electrical-, and electrochemical-coupled model for cylindrical
  wound large format lithium-ion batteries}, J. Power Sources, 241 (2013),
  pp.~20--32.

\bibitem{madani2018}
{\sc S.~Madani, E.~Schaltz, and S.~K{\ae}r}, {\em A review of different
  electric equivalent circuit models and parameter identification methods of
  lithium-ion batteries}, {ECS} Transactions, 87 (2018), pp.~23--37,
  \url{https://doi.org/10.1149/08701.0023ecst}.

\bibitem{marquis2019}
{\sc S.~Marquis, V.~pSulzer, R.~Timms, C.~Please, and S.~Chapman}, {\em An
  asymptotic derivation of a single article model with electrolyte}, J.
  Electrochem. Soc., 166 (2019), pp.~A3693--A3706.

\bibitem{marquis2020}
{\sc S.~Marquis, R.~Timms, V.~Sulzer, C.~Please, and S.~Chapman}, {\em A suite
  of reduced-order models of a single-layer lithium-ion pouch cell}, arXiv:
  2008.03691,  (2020).

\bibitem{McCleary2013}
{\sc D.~McCleary, J.~Meyers, and B.~Kim}, {\em {Three-Dimensional} modeling of
  electrochemical performance and heat generation of spirally and prismatically
  wound {Lithium-Ion} batteries}, J. Electrochem. Soc., 160 (2013),
  pp.~A1931--A1943.

\bibitem{moyles2018}
{\sc I.~Moyles, M.~Hennessy, T.~Myers, and B.~Wetton}, {\em Asymptotic
  reduction of a porous electrode model for lithium-ion batteries}, arXiv
  preprint arXiv:1805.07093,  (2018).

\bibitem{papanicolau1978}
{\sc G.~Papanicolau, A.~Bensoussan, and J.-L. Lions}, {\em Asymptotic analysis
  for periodic structures}, Elsevier, 1978.

\bibitem{pavliotis2008}
{\sc G.~Pavliotis and A.~Stuart}, {\em Multiscale methods, volume 53 of texts
  in applied mathematics}, 2008.

\bibitem{plett2015battery}
{\sc G.~Plett}, {\em Battery management systems, Volume I: Battery modeling},
  vol.~1, Artech House, 2015.

\bibitem{richardson2012}
{\sc G.~Richardson, G.~Denuault, and C.~Please}, {\em Multiscale modelling and
  analysis of lithium-ion battery charge and discharge}, J. Eng. Math., 72
  (2012), pp.~41--72.

\bibitem{richardson2019}
{\sc G.~Richardson, I.~Korotkin, R.~Ranom, M.~Castle, and J.~Foster}, {\em
  Generalised single particle models for high-rate operation of graded
  lithium-ion electrodes: Systematic derivation and validation}, Electrochimica
  Acta, 339 (2020), p.~135862,
  \url{https://doi.org/10.1016/j.electacta.2020.135862}.

\bibitem{rieger2016}
{\sc B.~Rieger, S.~Erhard, S.~Kosch, M.~Venator, A.~Rheinfeld, and A.~Jossen},
  {\em Multi-dimensional modeling of the influence of cell design on
  temperature, displacement and stress inhomogeneity in large-format
  lithium-ion cells}, J. Electrochem. Soc., 163 (2016), pp.~A3099--A3110.

\bibitem{sanchez1980}
{\sc E.~S{\'a}nchez-Palencia}, {\em Non-homogeneous media and vibration
  theory}, Lecture notes in physics, 127 (1980).

\bibitem{Saw2013}
{\sc L.~Saw, Y.~Ye, and A.~Tay}, {\em Electrochemical--thermal analysis of
  18650 lithium iron phosphate cell}, Energy Convers. Manage., 75 (2013),
  pp.~162--174.

\bibitem{scrosati2010}
{\sc B.~Scrosati and J.~Garche}, {\em Lithium batteries: Status, prospects and
  future}, J. Power Sources, 195 (2010), pp.~2419--2430.

\bibitem{Shi2018}
{\sc B.~Shi, H.~Zhang, Y.~Qi, and L.~Yang}, {\em Calculation model of effective
  thermal conductivity of a spiral-wound lithium ion battery}, J. Therm. Sci.,
  27 (2018), pp.~572--579.

\bibitem{timms2020}
{\sc R.~Timms, S.~Marquis, , V.~Sulzer, C.~Please, and S.~Chapman}, {\em
  Asymptotic reduction of a lithium-ion pouch cell model}, arXiv: 2005.05127,
  (2020).

\bibitem{tranter2020}
{\sc T.~Tranter, R.~Timms, T.~Heenan, S.~Marquis, V.~Sulzer, A.~Jnawali,
  M.~Kok, C.~Please, S.~Chapman, P.~Shearing, and D.~Brett}, {\em Probing
  heterogeneity in li-ion batteries with coupled multiscale models of
  electrochemistry and thermal transport using tomographic domains}, J.
  Electrochem. Soc., 167 (2020), p.~110538.

\bibitem{VanNoorden2014}
{\sc R.~Van~Noorden}, {\em The rechargeable revolution: A better battery},
  Nature News, 507 (2014), p.~26.

\bibitem{zubi2018}
{\sc G.~Zubi, R.~Dufo-L{\'o}pez, M.~Carvalho, and G.~Pasaoglu}, {\em The
  lithium-ion battery: State of the art and future perspectives}, Renew. Sust.
  Energ. Rev., 89 (2018), pp.~292--308.

\end{thebibliography}

\end{document}


\maketitle

\renewcommand{\thefootnote}{\fnsymbol{footnote}}
\footnotetext[2]{School of Mathematical Sciences, Queensland University of Technology, Brisbane QLD 4000, Australia and ARC Centre of Excellence for Mathematical and Statistical Frontiers, Queensland University of Technology, Brisbane QLD 4000, Australia (\email{steven.psaltis@qut.edu.au}).}
\footnotetext[3]{Mathematical Institute, Andrew Wiles Building, University of Oxford, Woodstock Road, Oxford OX2 6GG, UK (\email{timms@maths.ox.ac.uk}, \email{please@maths.ox.ac.uk}, \email{chapman@maths.ox.ac.uk}).}
\footnotetext[4]{The Faraday Institution, Quad One, Becquerel Avenue, Harwell Campus, Didcot, OX11 0RA, UK.}
\renewcommand{\thefootnote}{\arabic{footnote}}


\section{Reasonably conductive current collectors}
\label{sec:resonably conductive}
Here we  consider the intermediate limit in which the electrical conductivity of the current collector materials is reasonably large compared to that of the active material.
Specifically we consider the distinguished limit in which
$\sigma^{a_k} = \varepsilon^2 \bar{\sigma}^{a_k}$, where $\bar{\sigma}^{a_k}=O(1)$ as $\veps\ra 0$, for $k=1,2$.

The asymptotic analysis of this case is very similar to the very conductive case of Section 3.2. The main difference is that the behaviour in the active region comes into the analysis of the current collectors at second order, even though we still have to continue to fourth order in the current collectors to obtain the final solvability condition. Although this makes the algebra more complicated, it does not significantly alter the process.

The appropriate scaling of $\phi$ to maintain a correct balance with the source term is the same as in Section 3.2. Thus
   written in multiple-scales form equation (2.5) is
\begin{align}
\label{eqn:nondimmediumpm}  \lefteqn{\frac{\sigma^i}{r}\pd{}{r}\left(r\pd{\phi^i}{r}\right) + \frac{\sigma^i}{r^2}\spd{\phi^i}{\theta} + \frac{\sigma^i}{\varepsilon r}\pd{}{r}\left(r\pd{\phi^i}{R}\right) + \frac{\sigma^i}{\varepsilon r}\pd{}{R}\left(r\pd{\phi^i}{r}\right) + \frac{\sigma^i}{\varepsilon^2 }\spd{\phi}{R}}\hspace{4cm}&\\
  &= -\veps^2 F^i(r,R-\theta/2\pi),\nonumber \mbox{ }\quad i \in \{+, -\}
		,\\[2mm]
\label{eqn:nondimmediuma12} \lefteqn{\frac{\veps^2\bar{\sigma}^i}{r}\pd{}{r}\left(r\pd{\phi^i}{r}\right) + \frac{\veps^2\bar{\sigma}^i}{r^2}\spd{\phi^i}{\theta} + \frac{\veps\bar{\sigma}^i}{ r}\pd{}{r}\left(r\pd{\phi^i}{R}\right) + \frac{\veps\bar{\sigma}^i}{ r}\pd{}{R}\left(r\pd{\phi^i}{r}\right) + \bar{\sigma}^i\spd{\phi}{R}}\hspace{4.3cm}&\\
&= -\veps^2 F^i(r,R-\theta/2\pi), \qquad\qquad\non \mbox{ } \quad i \in \{ a_1, a_2\},
\end{align}
with the solution periodic in $R$ and $\theta$ with period $1$ and $2 \pi$ respectively. The continuity conditions (3.12)-(3.14) are unchanged, while
(3.15)-(3.17) become 
\begin{align}
\left.\sigma^+	\vec{n} \cdot \del \phi^+ \right|_{R = R_0 + \delta^+ + \pdhfrac{\theta}{2\pi}}&= 
	\left.	\varepsilon^2 \bar{\sigma}^{a_1}\vec{n} \cdot \del\phi^{a_1} \right|_{R = R_0 + \delta^+ + \pdhfrac{\theta}{2\pi}},
	\label{eq:conditions3a_r}\\
\left.\sigma^+	\vec{n} \cdot \del \phi^+\right|_{R = R_0  - \delta^+ + \pdhfrac{\theta}{2\pi}} &= 
	\left.	\varepsilon^2 \bar{\sigma}^{a_2}\vec{n} \cdot \del\phi^{a_2}\right|_{R = R_0 + 1 - \delta^+ + \pdhfrac{\theta}{2\pi}},
	\label{eq:conditions3b_r}\\
\sigma^-	\vec{n} \cdot \del\phi^- &= 
	\begin{cases}
		\varepsilon^2\bar{\sigma}^{a_1}\vec{n} \cdot \del\phi^{a_1} & \text{on} \  R = R_0 + \sep - \delta^- + \pdhfrac{\theta}{2\pi},\\
		\varepsilon^2 \bar{\sigma}^{a_2}\vec{n} \cdot \del\phi^{a_2} & \text{on} \  R = R_0 + \sep + \delta^- + \pdhfrac{\theta}{2\pi}.
	\end{cases}
	\label{eq:conditions4_r}
\end{align}
As usual we expand the solution in powers of $\varepsilon$ in the form
\begin{equation}
\phi^i \sim \phi_0^i + \varepsilon \phi_1^i + \varepsilon^2 \phi_2^i + \ldots,
\end{equation}
and equate coefficients of powers of $\varepsilon$.
Starting with the positive current collector, at leading and first order the analysis is identical to that in Section~3.2, so that
\begin{equation}
	\phi_0^+ = \phi_0^+(r, \theta),
\qquad
\phi_1^+ = -\pd{\phi_0^+}{r}\left(R - R_0 - \frac{\theta}{2\pi}\right) + \beta_1^+(r, \theta).
\label{eq:phi1p_r}
\end{equation}
At second order we again find 
\begin{equation}
    \phi_2^+ = \frac{1}{2}\left(-\frac{1}{r^2}\spd{\phi_0^+}{\theta} + \spd{\phi_0^+}{r}\right)\left(R - R_0 - \frac{\theta}{2 \pi}\right)^2 + \alpha_2^+\left(R - R_0 - \frac{\theta}{2 \pi}\right) + \beta_2^+,
    \label{eq:phi2sol_r}
\end{equation}
with $\alpha_2^+=\alpha_2^+(r, \theta)$, $\beta_2^+=\beta_2^+(r, \theta)$, 
but  (\ref{eq:conditions3a_r})-(\ref{eq:conditions4_r}) now give that
\begin{align}
  -\frac{1}{2\pi r^2}\pd{\phi_0^+}{\theta} - \frac{1}{r^2} \spd{\phi_0^+}{\theta}\delta^+ + \pd{\beta_1^+}{r} + \alpha_2^+ &= \frac{\bar{\sigma}^{a_1}}{\sigma^+}\left.\pd{\phi_0^{a_1}}{R}\right|_{R = R_0 + \delta^+ + \frac{\theta}{2 \pi}},\label{eq:bc2order1_r}\\
  -\frac{1}{2\pi r}\pd{\phi_0^+}{\theta} + \frac{1}{r^2} \spd{\phi_0^+}{\theta}\delta^+ + \pd{\beta_1^+}{r} + \alpha_2^+ & = \frac{\bar{\sigma}^{a_2}}{\sigma^+}\left.\pd{\phi_0^{a_2}}{R}\right|_{R = R_0 +1- \delta^+ + \frac{\theta}{2 \pi}}
	.\label{eq:bc2order2_r}
\end{align}
The leading-order potential in the active material this time  satisfies
\[ \spd{\phi_0^{a_k}}{R} = 0, \qquad k =1,2.\]
Continuity of the potential at the boundaries with the current collectors (3.12)-(3.14)   gives
\begin{align}
  \phi_0^{a_1} &= (\phi_0^--\phi_0^+) \frac{\left( R - R_0 - \delta ^+- \frac{\theta}{2 \pi} \right)}{\sep-\delta^+-\delta^-} + \phi_0^+,\label{pp-1b_r}
  \\
\phi_0^{a_2} &= (\phi_0^--\phi_0^+) \frac{\left( R_0 + 1 - \delta^+ + \frac{\theta}{2 \pi} - R\right)}{1-\sep - \delta^+-\delta^-} + \phi_0^+,
\label{pp-2b_r}
\end{align}
Subtracting \eref{eq:bc2order1_r} from \eref{eq:bc2order2_r} and using  \eref{pp-1b_r}-\eref{pp-2b_r}  we obtain  
\begin{equation}
  \frac{2\delta^+\sigma^+}{r^2}\PDD{\phi_0^+}{\theta}= -  \left(\frac{\bar{\sigma}^{a_1}}{\aone}+\frac{\bar{\sigma}^{a_2}}{\atwo} \right)(\phi_0^--\phi_0^+) .
    \label{eq:poscc_r}
\end{equation}
A similar analysis of the negative current collector yields
\begin{equation}
   \frac{2\delta^-\sigma^-}{r^2}\PDD{\phi_0^-}{\theta}= -  \left(\frac{\bar{\sigma}^{a_1}}{\aone}+\frac{\bar{\sigma}^{a_2}}{\atwo} \right)(\phi_0^+-\phi_0^-).
    \label{eq:negcc_r}
\end{equation}
The only solution of (\ref{eq:poscc_r})-(\ref{eq:negcc_r}) which is periodic in $\theta$ is $\phi_0^- = \phi_0^+ = \phi_0(r)$, for some function $\phi_0$, i.e. the leading-order potentials are equal and independent of $\theta$.
We can see this by subtracting an appropriate multiple of 
\eref{eq:poscc_r} from \eref{eq:negcc_r} to give
\begin{equation}
\frac{1}{r^2}\spd{}{\theta}(\phi_0^- - \phi_0^+) = \left(\frac{1}{\delta^+\sigma^+} +\frac{1}{\delta^-\sigma^-} \right)\left(\frac{\bar{\sigma}^{a_1}}{\aone}+\frac{\bar{\sigma}^{a_2}}{\atwo} \right) (\phi_0^- - \phi_0^+).
\end{equation}
Since, the coefficient of $(\phi_0^- - \phi_0^+)$ on the right-hand side is positive, periodicity in $\theta$
implies  \mbox{$\phi_0^- = \phi_0^+ = \phi_0$}, say, whence \eref{eq:poscc_r} implies
$\spd{\phi_0}{\theta} = 0$,
so that, again by periodicity, $\phi_0$ is independent of $\theta$.
Equations \eref{eq:bc2order1_r}-\eref{eq:bc2order2_r} then give
\begin{equation}
    \alpha_2^+=-\pd{\beta_1^+}{r}, \qquad \alpha_2^- = -\pd{\beta_1^-}{r},
\end{equation}
so that
\begin{equation}
	\phi_2^+ = \frac{1}{2}\left(\sdd{\phi_0}{r}\right)\left(R - R_0 - \frac{\theta}{2 \pi}\right)^2 - \pd{\beta_1^+}{r}\left(R - R_0 - \frac{\theta}{2 \pi}\right) + \beta_2^+.
	\label{eq:phi2_r}
\end{equation}
At the next order we find
\begin{align}
   \label{eq:phi3_r}    \phi_3^+ &= -\frac{1}{6}\tdd{\phi_0}{r}\left(R - R_0 - \frac{\theta}{2 \pi}\right)^3  +\frac{1}{2}\left(\PDD{\beta_1^+}{r}
    - \frac{1}{r^2} \PDD{\beta_1^+}{\theta}\right)\left(R - R_0 - \frac{\theta}{2 \pi}\right)^2\\
    &\mbox{ }\qquad \qquad + \alpha_3^+\left(R - R_0 - \frac{\theta}{2 \pi}\right) + \beta_3^+
\non ,
\end{align}
in the positive current collector and 
 \begin{align}
   \phi_1^{a_1} &= \alpha_1^{a_1} \frac{\left( R - R_0 - \delta^+ - \frac{\theta}{2 \pi} \right)}{\sep - \delta^+-\delta^-} + \beta_1^{a_1},& 
    \phi_1^{a_2} &= \alpha_1^{a_2} \frac{\left( R_0 + 1 - \delta^+ + \frac{\theta}{2 \pi} - R\right)}{1-\sep - \delta^+ - \delta^-} + \beta_1^{a_2},
 \end{align}
in the active regions $a_1$ and $a_2$,
Continuity at the boundaries
(3.12)-(3.14) gives
 \begin{align}
     \beta_1^{a_1} &= \beta_1^+ - \delta^+ \fdd{\phi_0}{r},&
     \beta_1^{a_2} &= \beta_1^+ + \delta^+ \fdd{\phi_0}{r},\\
     \alpha_1^{a_1} &= \beta_1^- - \beta_1^+ +(\delta^++\delta^-) \fdd{\phi_0}{r},&
     \alpha_1^{a_2} &= \beta_1^- - \beta_1^+ - (\delta^++\delta^-) \fdd{\phi_0}{r},
 \end{align}
 after which (3.15)-(3.17) give
 \begin{align}
   \sigma^+
   \left.\left(-\frac{1}{2 \pi r^2}\pd{\phi_1^+}{\theta} + \pd{\phi_2^+}{r} + \pd{\phi_3^+}{R}\right)\right|_{R = R_0 + \delta^+ + \pdhfrac{\theta}{2 \pi}} &= \bar{\sigma}^{a_1}\left( \frac{\alpha_1^{a_1}}
       {\sep - \delta^+-\delta^-}+\fdd{\phi_0}{r}\right),\label{eq:O4bc1_r}\\
       \sigma^+	\left.\left(-\frac{1}{2 \pi r^2}\fdd{\phi_1}{\theta} + \pd{\phi_2^+}{r} + \pd{\phi_3^+}{R}\right)\right|_{R = R_0 - \delta^+ + \pdhfrac{\theta}{2 \pi}} &=
       \bar{\sigma}^{a_2}\left( \frac{-\alpha_1^{a_2}}{1 -\sep - \delta^+-\delta^-}+\fdd{\phi_0}{r}\right)
\label{eq:O4bc2_r}.
 \end{align}
Subtracting \eref{eq:O4bc1_r} from \eref{eq:O4bc2_r}, using  \eref{eq:phi1p_r}, \eref{eq:phi2_r} and \eref{eq:phi3_r} gives
 \beqas
 \frac{2 \delta^+\sigma^+}{r^2} \spd{\beta_1^+}{\theta} & =&
 -\left(\frac{\bar{\sigma}^{a_1}}{\aone}+ \frac{\bar{\sigma}^{a_2}}{\atwo}\right)(\beta_1^- - \beta_1^+) -
 \frac{\bar{\sigma}^{a_1} \sep}{\aone}\fdd{\phi_0}{r}+
 \frac{\bar{\sigma}^{a_2} (1-\sep)}{\atwo}\fdd{\phi_0}{r}.
 \eeqas
 From the negative current collector we find
  \beqas
 \frac{2 \delta^-\sigma^-}{r^2} \spd{\beta_1^-}{\theta} & =&
 \left(\frac{\bar{\sigma}^{a_1}}{\aone}+ \frac{\bar{\sigma}^{a_2}}{\atwo}\right)(\beta_1^- - \beta_1^+) +
 \frac{\bar{\sigma}^{a_1} \sep}{\aone}\fdd{\phi_0}{r}-
 \frac{\bar{\sigma}^{a_2} (1-\sep)}{\atwo}\fdd{\phi_0}{r}.
 \eeqas
 Imposing periodicity now gives that
 \[ \left(\frac{\bar{\sigma}^{a_1}}{\aone}+ \frac{\bar{\sigma}^{a_2}}{\atwo}\right)(\beta_1^- - \beta_1^+) =-
 \frac{\bar{\sigma}^{a_1} \sep}{\aone}\fdd{\phi_0}{r}+
 \frac{\bar{\sigma}^{a_2} (1-\sep)}{\atwo}\fdd{\phi_0}{r},
 \]
 i.e.
 \[\beta_1^- - \beta_1^+ = -\mu \fdd{\phi_0}{r} + \left(\frac{\bar{\sigma}^{a_1}}{\aone}+ \frac{\bar{\sigma}^{a_2}}{\atwo}\right)^{-1}\frac{\bar{\sigma}^{a_2}}{\atwo}
 \fdd{\phi_0}{r},\]
 which then gives $\beta_1^+$, $\beta_1^-$ independent of $\theta$.
 The boundary conditions \eref{eq:O4bc1_r}-\eref{eq:O4bc2_r} then give
\beqas
  \alpha_3^+ &=& \left(\frac{1}{(\atwo/ \bar{\sigma}^{a_2} + \aone/ \bar{\sigma}^{a_1
    })\sigma^+} + \frac{1}{4\pi^2 r^2} \right)
  \fdd{\phi_0}{r} - \pd{\beta_2^+}{r},\\
  \alpha_3^- &=& \left(\frac{1}{(\atwo/ \bar{\sigma}^{a_2} + \aone/ \bar{\sigma}^{a_1
    })\sigma^-} + \frac{1}{4\pi^2 r^2} \right)
  \fdd{\phi_0}{r} - \pd{\beta_2^-}{r}.
\eeqas
At next order
\begin{align}
  \label{eq:phi4_r}\pd{\phi_4^+}{R} &= \frac{1}{6}
\phi_0''''
  \left(R-R_0-\frac{\theta}{2 \pi}\right)^3 - \frac{1}{2}\frac{\d^3\beta_1^+}{\d r^3}\left(R-R_0-\frac{\theta}{2 \pi}\right)^2 \\
&\qquad\mbox{ }- \left(
\left(2 \phi_0'' + \frac{\phi_0'}{r}\right)\frac{1}{(\atwo/ \bar{\sigma}^{a_2} + \aone/ \bar{\sigma}^{a_1
  })\sigma^+} \right.\nonumber\\
& \left.\qquad \qquad\qquad\mbox{ }+ \frac{3}{4 \pi^2 r} \left(\frac{\phi_0'}{r}\right)'
	+ \frac{1}{r^2}\spd{\beta_2^+}{\theta} 
	- \spd{\beta_2^+}{r}\right)\left(R-R_0-\frac{\theta}{2 \pi}\right)\nonumber\\ 
	&\qquad\mbox{ }+ \alpha_4^+
	 - \frac{1}{\sigma^+}\int_{R_0}^{R-\theta/2\pi} F^+(r,y)\, \d y,
\nonumber
\end{align}
where $' \equiv \d/\d r$, while
\begin{align*}
   \phi_2^{a_1} &= \left(-\frac{1}{\bar{\sigma}^{a_1}}\left(2\phi_0''+\frac{\phi_0'}{r}\right)\left(
\frac{1}{\atwo/\bar{\sigma}^{a_2}+ \aone/\bar{\sigma}^{a_1}}
\right)
-\phi_0''\right)\frac{(R-R_0-\theta/2\pi-\delta^+)^2}{2}\\
& \mbox{ }+  \alpha_2^{a_1}(R-R_0-\theta/2\pi-\delta^+) +\beta_2^{a_1}
   - \frac{1}{\bar{\sigma}^{a_1}}\int^{R-\theta/2\pi}_{R_0+\delta^+}(R-\theta/2\pi-y) F^{a_1}(r,y)\, \d y,  \\
   \phi_2^{a_2} &= \left(-\frac{1}{\bar{\sigma}^{a_2}}\left(2\phi_0''+\frac{\phi_0'}{r}\right)\left(
\frac{1}{\atwo/\bar{\sigma}^{a_2}+ \aone/\bar{\sigma}^{a_1}}
\right)
-\phi_0''\right)\frac{(R-R_0-\theta/2\pi-\sep-\delta^-)^2}{2}\\
& \!\!\!\!\!\!\mbox{}+ \alpha_2^{a_2} (R-R_0-{\theta}/{2\pi}-\sep-\delta^-) +\beta_2^{a_2}
   -  \frac{1}{\bar{\sigma}^{a_2}}\int^{R-\theta/2\pi}_{R_0+\sep+\delta^- } (R-\theta/2\pi-y)F^{a_2}(r,y)\, \d y.
\end{align*}
Following the same procedure of applying the continuity conditions (3.12)-(3.14) to determine $\alpha_2^{a_k}$, $\beta_2^{a_k}$, $k=1,2$, and subtracting the flux conditions  (\ref{eq:conditions3a_r})-(\ref{eq:conditions4_r}) gives
\beqas
 \frac{2 \delta^+\sigma^+}{r^2} \spd{\beta_2^+}{\theta} & =&
 -\left(\frac{\bar{\sigma}^{a_1}}{\aone}+ \frac{\bar{\sigma}^{a_2}}{\atwo}\right)(\beta_2^- - \beta_2^+) +f^+(r),
 \eeqas
 where $f^+$ is a long expression involving $\theta_0$ and $\beta_1^+$.
 From the negative current collector we find
  \beqas
 \frac{2 \delta^-\sigma^-}{r^2} \spd{\beta_2^-}{\theta} & =&
 \left(\frac{\bar{\sigma}^{a_1}}{\aone}+ \frac{\bar{\sigma}^{a_2}}{\atwo}\right)(\beta_2^- - \beta_2^+) +f^-(r),
 \eeqas
where $f^-$ is also a  long expression involving $\theta_0$ and $\beta_1^+$. The condition for periodicity is $f^++f^-=0$, which is the final equation for $\theta_0$ that we seek. The terms involving $\beta_1^+$ cancel, leaving, finally, after much simplification, 
 \begin{equation}
\frac{1}{(\aone/\bar{\sigma}^{a_1} + \atwo/\bar{\sigma}^{a_2})}	\frac{1}{r}\fdd{}{r}\left(  r\fdd{\phi_0}{r} \right)+	\frac{(\delta^+\sigma^+ +\delta^-\sigma^-)}{2 \pi^2r}\fdd{}{r}\left(  \frac{1}{ r}  \fdd{\phi_0}{r} \right)
	= -\bar{F}.
	\label{eq:phi_resonable}
\end{equation}

\subsection{Ohmic heating}
In the thermal problem the source term is given by Ohmic heating. The leading term in the current conductors is again from the azimuthal current, so that
\beqas
 \sigma^{\pm} | \del \phi|^2& \sim &
\frac{\veps^2\sigma^{\pm}}{r^2} \left( \pd{\phi_1^{\pm}}{\theta}
\right)^2 =\frac{\veps^2\sigma^{\pm}}{4 \pi^2 r^2} \left( \fdd{\phi_0}{r}
\right)^2 ,
\eeqas
while in the active material the Ohmic heating is
\beqas
 \veps^2 \bar{\sigma}^{a_k} | \del \phi|^2 &\sim& \veps^2 \bar{\sigma}^{a_k}  \left(\pd{\phi_1^{a_k}}{R}+\pd{\phi_0^{a_k}}{r}\right)^2 =
\frac{\veps^2}{\bar{\sigma}^{a_k} }\frac{1}{(\aone/\bar{\sigma}^{a_1} + \atwo/\bar{\sigma}^{a_2})^2}\left( \fdd{\phi_0}{r}
\right)^2  .
\eeqas

\section{Boundary conditions}
\label{sec:SM2}
\subsection{Boundary layer analysis near $r=r_0$}
The first turn of the winding behaves differently to the bulk, and this needs to be taken into account when formulating the boundary conditions. In the poor and very conductive cases (i) and (iii) the effect of the boundary layer is small, and it may be safely ignored. However, in the reasonably conductive problem (ii) its effect may be significant. Here we analyse this boundary layer to determine a more accurate  effective boundary condition that can be applied to the homogenised model.

\begin{figure}
  \begin{subfigure}[b]{0.45 \textwidth}
    \begin{overpic}[width=\textwidth,trim =-10mm -10mm 0 0]{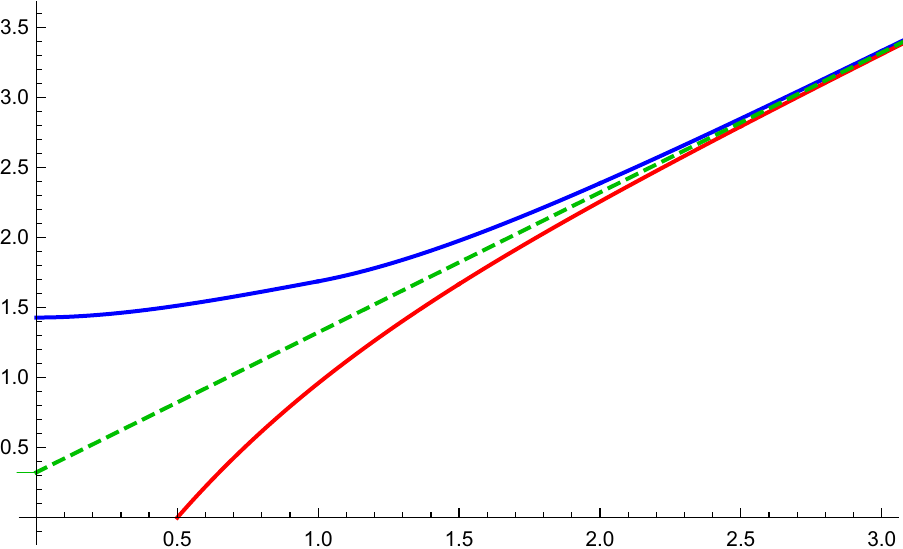}
      \put(40,39){$\phi^+$}
      \put(52,30){$\phi^-$}
      \put(0,17){$1/\alpha$}
      \put(50,5){$\rho$}
      \end{overpic}
  \end{subfigure}
  \qquad
 \begin{subfigure}[b]{0.45 \textwidth}
    \begin{overpic}[width=\textwidth,trim =0 -10mm 0 0]{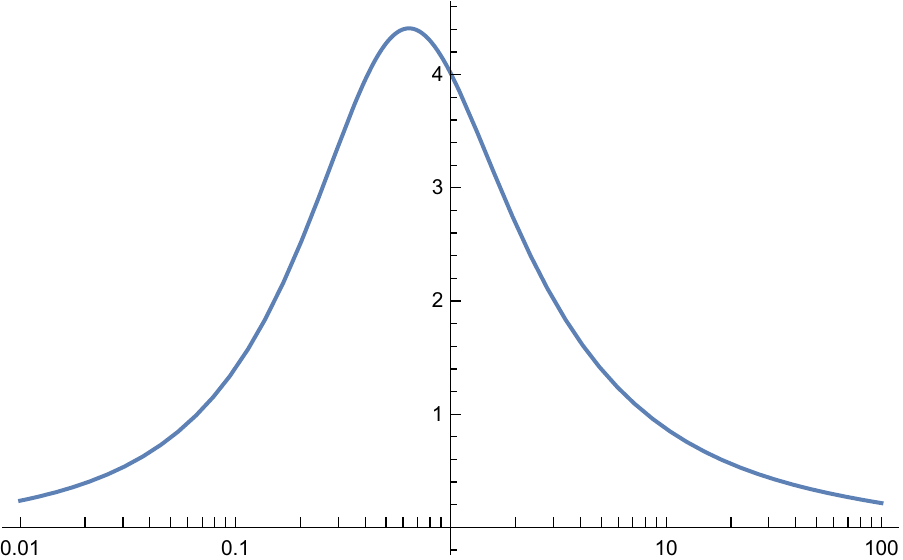}
      \put(52,70){$\alpha$}
       \put(50,5){$\omega$}
     \end{overpic}
 \end{subfigure}
 \caption{}
 \label{figSM1}
\end{figure}

We start from the following dimensionless composite model, valid in both the reasonable and very conductive regimes:
 \begin{align}
 \label{eq1}  \lefteqn{ \frac{\veps^2 \delta^+\sigma^+}{2 \pi^2} \frac{1}{r} \fdd{}{r}\left( \frac{1}{r} \fdd{\phi^+}{r}\right)+\frac{\sigma^{a_1}}{\aone \veps^2} \left( \frac{(r+\veps/2)}{r}\phi^-(r+\veps/2) - \phi^+(r)\right)}\hspace{10cm}&& \\
   \non \mbox{ }  + \frac{\sigma^{a_2}  }{\atwo \veps^2}\left( \frac{(r-\veps/2)}{r}\phi^-(r-\veps/2) - \phi^+(r)\right)
&=
 0,\\
 \label{eq2}\lefteqn{  \frac{\veps^2 \delta^-\sigma^-}{2 \pi^2} \frac{1}{r} \fdd{}{r}\left( \frac{1}{r} \fdd{\phi^-}{r}\right)+ \frac{\sigma^{a_2}  }{\atwo \veps^2} \left( \frac{(r+\veps/2)}{r}\phi^+(r+\veps/2) - \phi^-(r)\right)}\hspace{10cm} &&\\
 \non \mbox{ }+ \frac{\sigma^{a_1}}{\aone \veps^2}\left( \frac{(r-\veps/2)}{r}\phi^+(r-\veps/2) - \phi^-(r)\right) 
  & = 0.
\end{align}
In the bulk $\phi^-$ and $\phi^+$ vary slowly with $r$ and terms evaluated at $r \pm \veps/2$ can be Taylor expanded resulting in the composite model (4.21)-(4.22).
However, $\phi^-(r)$ is only defined for $r\geq r_0+\veps/2$, so that equation (\ref{eq2}) is only valid for $r\geq r_0+\veps/2$, while 
in the first winding $r_0\leq r<r_0+\veps$ (\ref{eq1}) is replaced by
\begin{align*}
  \frac{\veps^2\delta^+\sigma^+}{2 \pi^2} \frac{1}{r} \fdd{}{r}\left( \frac{1}{r} \fdd{\phi^+}{r}\right)+\frac{\sigma^{a_1} (\phi^-(r+\veps/2) - \phi^+(r)) }{\aone \veps^2} 
&=
 0.
\end{align*}
The appropriate boundary conditions are
\[ \fdd{\phi^+}{r}(r_0) = 0, \qquad \phi^-(r_0+\veps/2) = 0.
\]
We simplify the discussion and reduce the number of parameters by supposing that $\sigma^{+} = \sigma^-$, $\sigma^{a_1} = \sigma^{a_2} = \sigma/\sigma^+$,  $\delta^+ = \delta^- = \delta$, $\aone=\atwo=\ell$. 
Then, defining $\hat{\phi}^-(r) = \phi^-(r+\veps/2)$, introducing the local  variable $\rho$ through $r = r_0+\veps \rho$,  we find that to  leading order in $\veps$ in the boundary layer we have
\begin{align}
\omega  \sdd{\phi^+}{\rho}
  +\hat{\phi}^-(\rho) - \phi^+(\rho)
&=
 0\qquad 0<\rho<1,\label{1a}\\
\omega \sdd{\phi^+}{\rho} +
 \hat{\phi}^-(\rho) - 2\phi^+(\rho)  +\hat{\phi}^-(\rho-1)  
&=
 0\qquad 1<\rho<\infty,\label{1b}\\
\omega \sdd{\hat{\phi}^-}{\rho}
 +\phi^+(\rho+1) - 2\hat{\phi}^-(\rho) +\phi^+(\rho)
  & = 0\qquad 0<\rho<\infty,\label{2}
\end{align}
with $\phi^+$, $\hat{\phi}^-$ and their derivatives continuous, and 
\beq \fdd{\phi^+}{\rho}(0) = 0, \qquad \hat{\phi}^-(0) = 0,\label{3}
\eeq
where
\[ \omega =  \frac{\delta\ell \veps^2}{2 \pi^2\sigma }\frac{1}{r_0^2} .\]
Note that $\omega = O(1)$ in the reasonably conductive regime.
To match with the bulk homogenised solution we require 
\beq
\phi^+ \sim \veps \fdd{\phi_R}{r}(r_0)\rho +\phi_R(r_0) , \quad \hat{\phi}^- \sim \veps \fdd{\phi_R}{r}(r_0)\left( \rho + \frac{1}{2}\right)+\phi_R(r_0) \qquad \mbox{ as } \rho \ra \infty.
\eeq
Since the problem is linear we solve (\ref{1a})-(\ref{2}) numerically with
\[ \phi^+ \sim \rho +\frac{1}{\alpha} , \quad \hat{\phi}^- \sim  \rho + \frac{1}{2} +\frac{1}{\alpha} \qquad \mbox{ as } \rho \ra \infty,\]
and determine the value of $\alpha$. A representative solution is shown in Fig.~\ref{figSM1}(a), and $\alpha$ is shown as a function of $\omega$ in Fig.~\ref{figSM1}(b). The appropriate boundary condition on the outer homogenised $\phi_R$ is then
\beq
\fdd{\phi_R}{r}(r_0) = \frac{\alpha}{\veps} \phi_R(r_0), \label{bcnew}
\eeq
or, for the dimensional model,
\beq
\fdd{\phi_R}{r}(r_0) = \frac{\alpha}{h} \phi_R(r_0).
\eeq
Formally of course at leading order (\ref{bcnew}) reduces to simply $\phi_R(r_0)=0$ as expected. However, if we instead impose (\ref{bcnew}) this accurately captures the $O(\veps)$ correction generated by the boundary layer .

For our numerical example with $\delta=0.05$, $\ell=0.4$, $\veps = 0.0375$, $r_0 =0.25$, $\sigma = 0.01 \veps^2$ we find $\omega \approx 1.62$ and $\alpha/\veps \approx 82.37$.

To provide an effective boundary condition for our composite model $\phi_{RV}$ we need to determine how it behaves in the boundary layer.
Locally the model RV satisfies
\begin{align}
\omega_1 \sdd{\phi^+}{\rho} +
 2{\phi}^-(\rho) - 2\phi^+(\rho) 
&=
 0\qquad 0<\rho<\infty,\label{RV1}\\
\omega_1 \sdd{{\phi}^-}{\rho}
 +2\phi^+(\rho) - 2{\phi}^-(\rho)   & = 0\qquad 0<\rho<\infty,\label{RV2}
\end{align}
with solution
\beqas
\phi^+ & = & A \rho + B + C \ee^{-2\rho/\sqrt{\omega_1}},\\
\phi^- & = & A \rho + B - C \ee^{-2\rho/\sqrt{\omega_1}},
\eeqas
where
\[ \omega_1 =  \frac{\delta\ell h^2}{2 \pi^2\sigma }\frac{1}{r_0^2}
+\frac{1}{4}  .\]
To match the far field of the boundary layer problem
 (\ref{1a})-(\ref{2})
we require $B = A/\alpha$.
Suppose we fix $\phi^-(1/2) = 0$, so that
\[  C =A\left(\frac{1}{2} + \frac{1}{\alpha}\right)\ee^{1/\sqrt{\omega_1}}.\]
Since there is just one free parameter remaining in the local solution, there is some freedom in writing the two effective boundary conditions on the outer homogenised solution. We have
\beqas
\fdd{\phi^+}{\rho}(0) & = & A\left(1 - \frac{2}{\sqrt{\omega_1}}\left(\frac{1}{2} + \frac{1}{\alpha}\right)\ee^{1/\sqrt{\omega_1}}\right),\\
\fdd{\phi^-}{\rho}(0) & = & A\left(1 +
\frac{2}{\sqrt{\omega_1}}\left(\frac{1}{2} + \frac{1}{\alpha}\right)\ee^{1/\sqrt{\omega_1}}\right),\\
\phi^+(0) & = & \frac{A}{\alpha} + A\left(\frac{1}{2} + \frac{1}{\alpha}\right)\ee^{1/\sqrt{\omega_1}},\\
\phi^-(0) & = & \frac{A}{\alpha} - A\left(\frac{1}{2} + \frac{1}{\alpha}\right)\ee^{1/\sqrt{\omega_1}}.
\eeqas
One option in eliminating $A$ is to impose
\beqas
\fdd{\phi^+}{\rho} & = & (\phi^+ + \phi^-)\frac{\alpha}{2}\left(1 - \frac{2}{\sqrt{\omega_1}}\left(\frac{1}{2} + \frac{1}{\alpha}\right)\ee^{1/\sqrt{\omega_1}}\right),\\
\fdd{\phi^-}{\rho} & = & (\phi^+ + \phi^-)\frac{\alpha}{2}\left(1 + \frac{2}{\sqrt{\omega_1}}\left(\frac{1}{2} + \frac{1}{\alpha}\right)\ee^{1/\sqrt{\omega_1}}\right),
\eeqas
which, in our example gives
\[
\fdd{\phi^+}{r}  = -61.85 (\phi^+ + \phi^-), \qquad
\fdd{\phi^-}{r}  =  144.22(\phi^+ + \phi^-),
\]
at $r=r_0$.

\subsection{Boundary layer analysis near $r=1$}
A similar analysis holds near the outer winding.
We subtract the constant potential, and again shift the point of evaluation of $\phi^-$, by writing $\phi^+(r) = 1+ \bar{\phi}^+(r)$, $\phi^-(r+\veps/2) = 1+\hat{\bar{\phi}}(r)$. Introducing  $r = 1 + \veps \rho$ we find to leading order in $\veps$ that 
\begin{align}
\omega \sdd{\hat{\bar{\phi}}^-}{\rho}
  - \hat{\bar{\phi}}^-(\rho) +\bar{\phi}^+(\rho)
  & = 0\qquad -1<\rho<0,\label{1a-1}\\
\omega \sdd{\bar{\phi}^+}{\rho} +
 \hat{\bar{\phi}}^-(\rho) - 2\bar{\phi}^+(\rho)  +\hat{\bar{\phi}}^-(\rho-1)  
&=
 0\qquad -\infty<\rho<0,\label{1b-1}\\
\omega \sdd{\hat{\bar{\phi}}^-}{\rho}
 +\bar{\phi}^+(\rho+1) - 2\hat{\bar{\phi}}^-(\rho) +\bar{\phi}^+(\rho)
  & = 0\qquad -\infty<\rho<-1,\label{2-1}
\end{align}
with
\beq \bar{\phi}^+(0) = 1, \qquad \fdd{\hat{\bar{\phi}}^-}{\rho}(0) = 0,\label{3-1}
\eeq
and
\beq
\bar{\phi}^+ \sim \veps \fdd{\phi_R}{r}(1) \rho + \phi_R(1)-1, \quad \hat{\bar{\phi}}^- \sim \veps \fdd{\phi_R}{r}(1)\left( \rho + \frac{1}{2}\right)+ \phi_R(1)-1 \qquad \mbox{ as } \rho \ra- \infty,
\eeq
where this time
\[ \omega =  \frac{\delta\ell \veps^2}{2 \pi^2\sigma } .\]
Solving  (\ref{1a-1})-(\ref{2-1}) numerically with
\[ \bar{\phi}^+ \sim \rho -\frac{1}{\beta} , \quad\hat{\bar{\phi}}^- \sim  \rho + \frac{1}{2} -\frac{1}{\beta} \qquad \mbox{ as } \rho \ra -\infty,\]
gives $\beta$ as a function of $\omega$ shown in Fig.~\ref{figSM1}(b);
a representative solution is shown in Fig.~\ref{figSM1}(a). The appropriate boundary condition on the outer homogenised $\phi_R$ is then
\beq
\fdd{\phi_R}{r}(1) = \frac{\beta}{\veps}(1- \phi_R(1)),
\eeq
or dimensionally,
\beq \fdd{\phi_R}{r}(1) = \frac{\beta}{h}(1- \phi_R(1)).
\eeq
For our numerical example with $\delta=0.05$, $\ell=0.4$, $\veps = 0.0375$, $r_0 =0.25$, $\sigma = 0.01 \veps^2$ we find $\omega \approx 0.101321$ and $\beta/\veps \approx 22.1616$.

\begin{figure}
  \begin{subfigure}[b]{0.45 \textwidth}
    \begin{overpic}[width=\textwidth,trim =0 -2mm 0 -10mm]{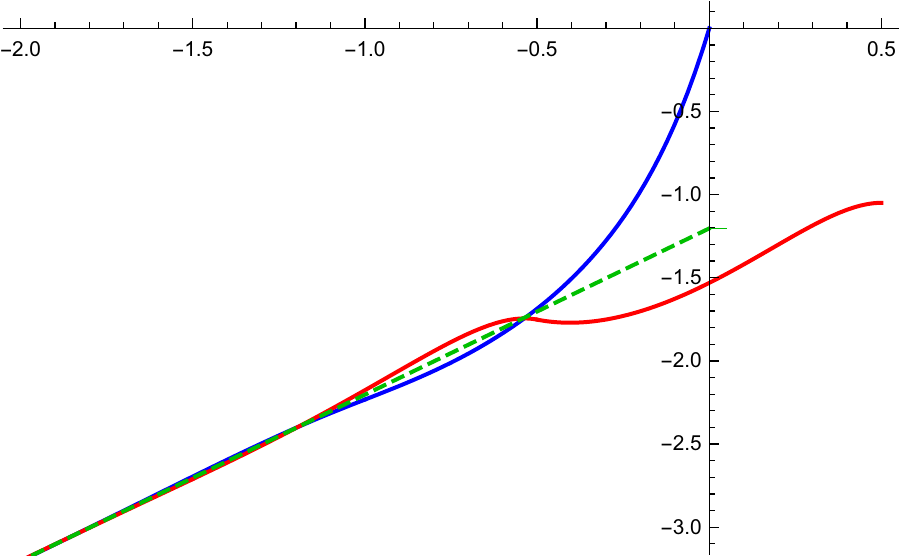}
      \put(70,53){$\phi^+$}
      \put(90,34){$\phi^-$}
      \put(80,39){$\scriptsize -1/\beta$}
      \put(50,65){$\rho$}
      \end{overpic}
  \end{subfigure}
  \qquad
  \begin{subfigure}[b]{0.45 \textwidth}
    \begin{overpic}[width=\textwidth,trim =0 -10mm 0 0]{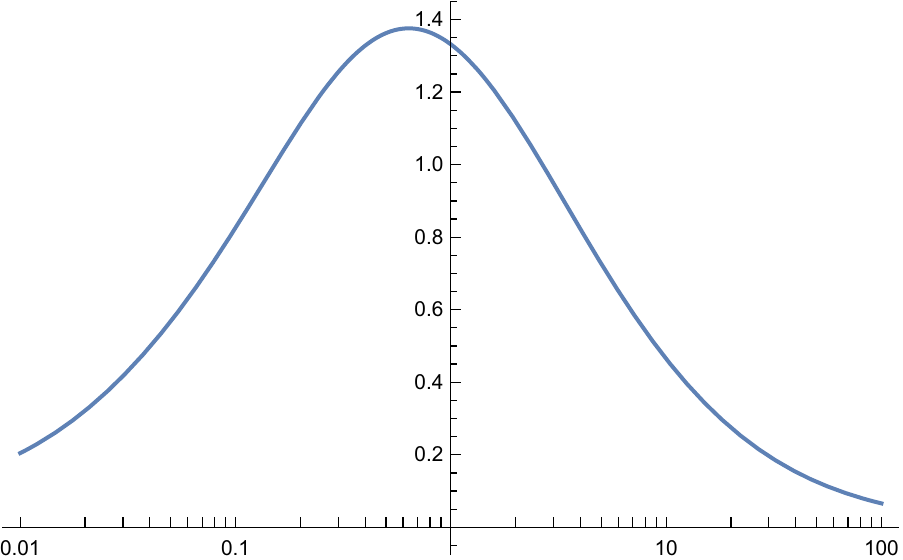}
      \put(52,70){$\beta$}
       \put(50,5){$\omega$}
     \end{overpic}
  \end{subfigure}
\caption{}   
\end{figure}
For the RV model we have locally 
\begin{align}
\omega_1 \sdd{\phi^+}{\rho} +
 2{\phi}^-(\rho) - 2\phi^+(\rho) 
&=
 0\qquad -\infty<\rho<0,\label{RV1-1}\\
\omega_1 \sdd{{\phi}^-}{\rho}
 +2\phi^+(\rho) - 2{\phi}^-(\rho)   & = 0\qquad -\infty<\rho<0,\label{RV2-1}
\end{align}
where this time
\[ \omega_1 =  \frac{\delta\ell h^2}{2 \pi^2\sigma }
+\frac{1}{4},\]
with solution
\beqas
\phi^+ & = & 1+ A \rho + B + C \ee^{2\rho/\sqrt{\omega_1}},\\
\phi^- & = & 1+ A \rho + B - C \ee^{2\rho/\sqrt{\omega_1}}.
\eeqas
To get the right behaviour in the far field we need $B =- A/\alpha$.
If  we fix $\phi^+(0) = 1$ then
\[  C =\frac{A}{\alpha},\]
so that
\begin{align*}
\fdd{\phi^+}{\rho}(0)  &=  A\left(1 + \frac{2}{\sqrt{\omega_1}}\frac{1}{\alpha}\right), &
\fdd{\phi^-}{\rho}(0)  &=  A\left(1 - \frac{2}{\sqrt{\omega_1}} \frac{1}{\alpha}\right),\\
\phi^+(0) & = 1, &
\phi^-(0) & = 1- \frac{2A}{\alpha}.
\end{align*}
As before there is some freedom in eliminating $A$ to give two boundary conditions.
One option is to impose
\beqas
\phi^+(0) & = & 1,\\
\fdd{\phi^-}{\rho}(0) & = &\frac{\alpha}{2}\left(1 - \frac{2}{\sqrt{\omega_1}}
 \frac{1}{\alpha}\right)(1-\phi^-(0)).
\eeqas
In our example this gives
\[
\phi^+  =  1, \qquad
\fdd{\phi^-}{r}  = -33.9093(1-\phi^-),
\]
at $r=1$.
